\documentclass[prx,english,superscriptaddress,floatfix,longbibliography,twocolumn]{revtex4-2}

\usepackage{tikz}
\usetikzlibrary{shapes, positioning, arrows, positioning, shadows}

\usepackage{colortbl}
\usepackage{siunitx}
\usepackage{graphicx}
\usepackage[T1]{fontenc}%
\usepackage[utf8]{inputenc}%
\usepackage{lmodern}%
\usepackage{textcomp}%
\usepackage{lastpage}%
\usepackage{geometry}%
\usepackage{amsmath}
\usepackage{amssymb}

\usepackage{listings}
\usepackage{longtable}
\usepackage{xfrac}

\usepackage{xcolor}
\usepackage[colorlinks,allcolors=blue]{hyperref}

\newcommand\ket[1]{\ensuremath{|#1\rangle}}
\newcommand\bra[1]{\ensuremath{\langle#1|}}

\newcommand{\tr}{\operatorname{tr}}

\newcommand{\ler}{p_{\mathrm{logical}}}
\newcommand{\hamparam}{\vec{\theta}_\mathrm{ham}}

\newcommand{\opn}{\hat{n}}
\newcommand{\opv}{\hat{\varphi}}

\newcommand{\aref}[1]{\hyperref[#1]{Appendix~\ref*{#1}}}

\newtheorem{assumption}{Assumption}

\def\<{\langle}
\def\>{\rangle}

\begin{document}

\title{Superconducting processor design optimization for quantum error correction performance}

\date{\today}

\author{Xiaotong Ni}
\email{xiaotong.ni@gmail.com}
\affiliation{Quantum Laboratory, DAMO Academy, Hangzhou, Zhejiang 311121, P.R.China}

\author{Ziang Wang}
\affiliation{Zhejiang Institute of Modern Physics, Zhejiang University, Hangzhou 310027, China}
\affiliation{Quantum Laboratory, DAMO Academy, Hangzhou, Zhejiang 311121, P.R.China}

\author{Rui Chao}
\affiliation{Quantum Laboratory, DAMO Academy, Bellevue, Washington 98004, USA}

\author{Jianxin Chen}
\email{jianxinchen@acm.org}
\affiliation{Quantum Laboratory, DAMO Academy, Bellevue, Washington 98004, USA}

\begin{abstract}

In the quest for fault-tolerant quantum computation using superconducting processors, accurate performance assessment and continuous design optimization stands at the forefront.
To facilitate both meticulous simulation and streamlined design optimization, we introduce a multi-level simulation framework that spans both Hamiltonian and quantum error correction levels, and is equipped with the capability to compute gradients efficiently. This toolset aids in design optimization, tailored to specific objectives like quantum memory performance. Within our framework, we investigate the often-neglected spatially correlated unitary errors, highlighting their significant impact on logical error rates. We exemplify our approach through the multi-path coupling scheme of fluxonium qubits.

\end{abstract}

\maketitle

\tableofcontents

\section{Introduction}

Fault-tolerant quantum computing (FTQC) remains the most promising and reliable approach to achieving practical quantum advantage, with a solid theoretical foundation.
Superconducting circuits have become a promising platform due to their low error rates and scalability. One pivotal step towards FTQC using these circuits is the creation of quantum memory with error correction, the cornerstone for digital quantum computers.
Despite recent advancements, such as the work by \cite{acharyaSuppressingQuantumErrors2023, zhaoRealizationErrorCorrectingSurface2022, krinnerRealizingRepeatedQuantum2022}, achieving error rates below the fault-tolerance threshold in processors of significant sizes remains elusive.
In order to progress towards high-quality quantum memory, it is imperative to focus on further refinement in chip design and advancements in quantum error correction (QEC) schemes.
Among the various fault-tolerant protocols proposed, the surface code~\cite{dennisTopologicalQuantumMemory2002, fowlerSurfaceCodesPractical2012} and its variants have emerged as extensively researched approaches that impose minimal hardware requirements on superconducting processors.
Consequently, we will utilize a surface code quantum memory as an example to illustrate our workflow. However, it is important to note that the workflow can be expanded to optimize other fault-tolerant schemes.

Indeed, even within the restricted scope of improving surface code quantum memory in superconducting processors, the design space remains vast.
Each qubit and the coupling between qubits involve multiple parameters that can be optimized.
Additionally, there are freedoms in how we control a processor to run the syndrome extraction process.
For instance, gate implementations can be engineered to convert amplitude damping errors into erasure errors, as demonstrated by \cite{levineDemonstratingLongcoherenceDualrail2023, chouDemonstratingSuperconductingDualrail2023, kubicaErasureQubitsOvercoming2022}.
The syndrome extraction circuits can also be modified to have lower qubit connectivity or native gates, as explored by \cite{mcewenRelaxingHardwareRequirements2023}.
However, conducting experiments to explore and optimize these parameters and schemes is resource-intensive and time-consuming.
Building and testing a large-scale processor with incremental parameter changes can take weeks, while developing and a new scheme and validating its scalability can require years.
Relying solely on experimental testing would result in excessively long duration to find optimal designs.
Therefore, the ability to evaluate quantum memory performance using numerical simulations is highly desirable.

However, challenges persist in accurately evaluating the performance of quantum memory from the design and control schemes of quantum processors.
This is still the case even when we start from the Hamiltonian descriptions of quantum processors, which we adopt as our starting point.
The exact simulation of time evolution for processors with a few dozen qubits is computationally infeasible.
The common approach to tackle this issue involves dividing time evolution into basic operations and approximating the error of each operation using stochastic Pauli errors.
However, there are two issues with this approach that need to be addressed.
\begin{enumerate}
\item Firstly, when dividing time evolution into basic operations, there exist gauge freedoms that can impact the simulated memory performance.
These gauge freedoms need to be carefully considered, and ideally we want to compute the logical error rates in a gauge-independent manner.
\item Secondly, and perhaps more importantly, errors occurring across multiple operations can be correlated.
Correlated errors are more detrimental to QEC performance and conflict with the strategy of examining one operation at a time.
This is an important aspect that needs to be taken into account in the evaluation process.
\end{enumerate}
In this work, we investigate a pervasive source of unitary errors with spatial correlation in superconducting processors, arising from the always-on interactions between qubits in the Hamiltonian.
Subsequently, we assess the influence of these errors on the logical error rates of memory operations.
These errors become increasingly significant as the rates of decoherence errors decrease, which is a common regime for comparing fault-tolerant schemes.
To mitigate such correlated errors, tunable couplers and related designs~\cite{gellerTunableCouplerSuperconducting2015,mundadaSuppressionQubitCrosstalk2019, yanTunableCouplingScheme2018, xuHighFidelityHighScalabilityTwoQubit2020,stehlikTunableCouplingArchitecture2021, collodoImplementationConditionalPhase2020,nguyenBlueprintHighPerformanceFluxonium2022, dingHighFidelityFrequencyFlexibleTwoQubit2023} prove to be effective.
For instance, it is shown that $ZZ$-crosstalk when qubits are idling can be made very small in~\cite{xuHighFidelityHighScalabilityTwoQubit2020,stehlikTunableCouplingArchitecture2021,nguyenBlueprintHighPerformanceFluxonium2022, dingHighFidelityFrequencyFlexibleTwoQubit2023}.
Nevertheless, our analysis demonstrates that the remaining correlated errors after mitigation cannot always be disregarded.
In particular, in~\autoref{sec:example_design_control_schemes}, we will use the fluxonium with multipath couplings~\cite{nguyenBlueprintHighPerformanceFluxonium2022} as an example to demonstrate our QEC simulation workflow.
We show that the remaining correlated errors in this scheme can increase logical error rates by roughly 5 times under practical parameters.
Moreover, when considering the implementation of low-density parity-check (LDPC) codes on superconducting processors, the impact of correlated errors can become even more substantial due to the higher connectivity of qubits.
Hence, it is crucial to incorporate both Hamiltonian-level and QEC-level simulations to provide a more accurate assessment of logical operation performance.

To improve the speed of evaluating and optimizing processor design and control schemes, we have developed an efficient method within our multi-level simulation framework to estimate the gradients of logical error rates of quantum memory and potentially other objective functions.
Previous works, such as computing gradients in optimal control~\cite{khanejaOptimalControlCoupled2005,leungSpeedupQuantumOptimal2017} or in device optimization~\cite{niIntegratingQuantumProcessor2022}, have primarily focused on the performance of individual quantum operations.
Therefore, our gradient computation method can be viewed as an extension of these previous works by considering the combined performance of various operations.
We showcase two optimization tasks in this paper.
\begin{enumerate}
    \item To compute the unitary errors of simultaneous gate operations (e.g.
see~\autoref{fig:simultaneous_1q_correlated_error}), we need to optimize control parameters such as drive frequencies and gate durations.
This is done with gradient optimization.
    \item We can set the objective function to be the logical error rate, and optimize the control parameters together with device parameters of superconducting qubits.
The gradients can be computed approximatedly and used in the optimization.
\end{enumerate}
By leveraging these gradient estimations, we can speedup the evaluation and optimization of processor design and control schemes.
Given the potential large number of parameters involved, such as over 50 in our optimization process, the utilization of gradient information leads to significant speed improvements.
Moreover, the obtained gradients can find applications in diverse contexts.
For instance, it can be utilized to estimate device parameters in the Hamiltonian description using experimental data.
One approach to accomplish this is by passing the computed gradient to the procedure outlined in~\cite{krastanovStochasticEstimationDynamical2019}.

This paper is organized as follows.
In Section~\ref{sec:theoretical_foundation}, we will examine several topics that are necessary for the simulation in the processor design workflow.
In Section~\ref{sec:workflow}, we propose a workflow for evaluating the logical error rates of the surface code quantum memory from the Hamiltonian description of the system.
This workflow approximates the strength of correlated errors and incorporates them into the simulation of syndrome extraction circuits.
In Section~\ref{sec:example_design_control_schemes}, we consider fluxonium qubits with multi-path coupling as an example to showcase the effectiveness of the proposed workflow.
Additionally, we demonstrate how gradient information can be utilized to optimize the parameters of simultaneous gate operations, as well as the parameters of fluxonium qubits and their couplings.

\section{Theoretical foundation}
\label{sec:theoretical_foundation}
In this section, we will present prior findings as well as new observations into the simulation of QEC tasks for superconducting processors.
These insights will inform the workflow described in~\autoref{sec:workflow}.
Specifically, we will detail common methods to turn off qubit interactions and elucidate why certain straying many-body effective interactions remain.
Additionally, we will shed light on the theoretical implications of these high-weight errors on QEC performance.

\subsection{Surface code and superconducting processors}

\begin{figure*}
    \includegraphics[width=0.8\linewidth]{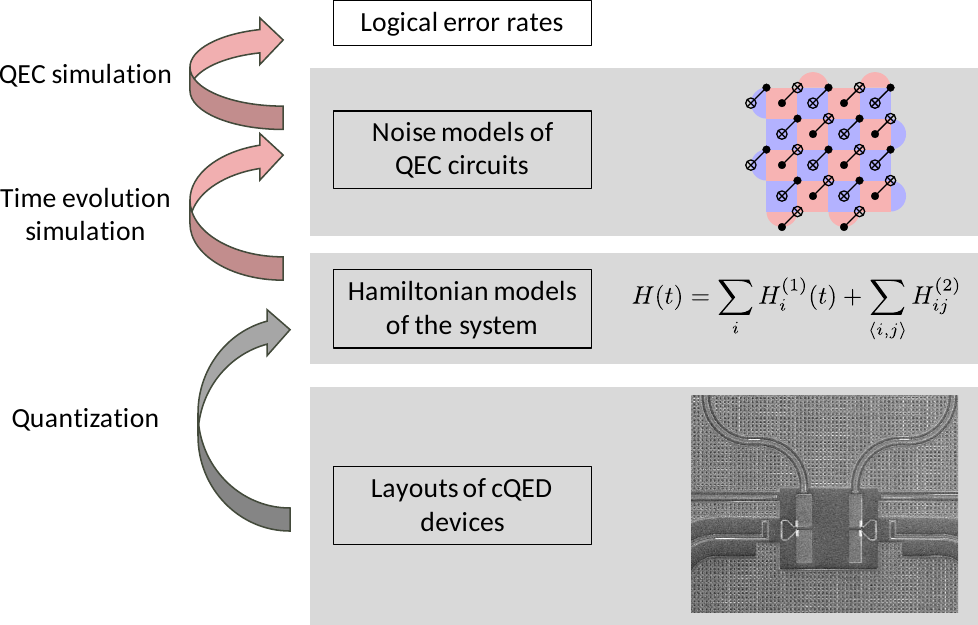}
    \caption{An overview of the various layers involved in conducting quantum memory experiments on superconducting processors is presented. Numerical simulations can be conducted from the bottom up, where parameters in the upper layers may be influenced by the simulation results of the underlying layers. Each layer may introduce certain approximations to keep the simulation manageable while still aiming for useful and accurate insights. This work contains time evolution and QEC simulations indicated by pink arrows. A more detailed flowchart of our simulations can be found in~\autoref{fig:workflow}.
    Descriptions of the Hamiltonian are provided in~\autoref{eq:2local_hamiltonian} and~\autoref{eq_hamiltonian_parameter}. The image which showcases the superconducting processor is from~\cite{baoFluxoniumAlternativeQubit2022}.
\label{fig:overall_cqed_qec}}
\end{figure*}

In the field of quantum error correction, various codes, such as the surface code, make use of analogies to classical error correction techniques. One of the main strategies involves encoding a set of $k$ qubits into a larger set of $n$ qubits, in order to introduce redundancies.
Among the prevailing choices for quantum error-correcting codes, stabilizer codes stand out. Extracting error information without disrupting the stored quantum data mandates the measurement of multi-qubit Pauli operators' eigenvalues, drawing a parallel to the parity checks in classical codes.
A pivotal parameter in QEC codes is the \textit{code distance} denoted by $d$ (presumed to be odd), which represents the fewest number of qubits one must act upon to execute a logical operation. Consequently, there are errors consisting of $\frac{d+1}{2}$ single-qubit Pauli operators can not be recovered perfectly.
Based on this, estimations have been made~\cite{fowlerSurfaceCodesPractical2012,watsonLogicalErrorRate2014} suggesting that when the independent error rate $p$ on each qubit is small enough, the logical error rate $e_L$ approximates to $O(f(d) p^{(d+1)/2})$, with $f(d)$ being a combinatorial factor. However, if each error event affects $k$ qubits (designated as weight-
$k$ errors) and such events occur at an independent error rate $p'$ , then the term
$p^{(d+1)/2}$ from the aforementioned estimation is replaced by $(p')^{\lfloor \frac{d}{2k} \rfloor +1}$.
It is evident that error weights profoundly influence logical error rates. A central objective of this work is to delve into the error locality in superconducting quantum processors.

The aforementioned offers a concise overview of quantum error correction. In most superconducting qubit architectures, multi-qubit stabilizer operators' eigenvalues can be measured by employing the syndrome extraction circuits, with one such implementation depicted in~\autoref{fig:syndrome_circuit}.
It is noteworthy, however, that since gates and measurements are inherently prone to faults, more errors might be introduced than what our selected QEC code can rectify. To gain insights into this practical scenario, simplified error models such as SD6 (standard depolarizing 6-step cycle)~\cite{gidneyFaultTolerantHoneycombMemory2021} are considered.
These models postulate a uniform probability $p$ for each gate and measurement to induce a random Pauli error on their respective acted-upon qubits. Under this model, the surface code exhibits a threshold of $p_\mathrm{th} \approx 0.57\%$, given the deployment of the minimum-weight perfect matching decoder~\cite{fowlerSurfaceCodesPractical2012}. When $p<p_\mathrm{th}$, the logical error rates diminish exponentially in relation to the code distance.
Yet, in reality, the errors of operations differ in their probabilities and localities.
For example, the non-uniform of gate errors on different qubits could stem from differences in coherence times, the parameters of the qubits, and more.
Additionally, it was observed in multiple experiments~\cite{zajacSpectatorErrorsTunable2021,chenExponentialSuppressionBit2021} that isolated versus simultaneous randomized benchmarking yield contrasting average gate errors.
Therefore, actual systems diverge considerably from the aforementioned simplified noise model SD6.
One consequence is that the fidelity of a singular two-qubit gate is not enough for deciding whether we have reached the threshold.

Currently, it is popular to consider quantum codes where each subsystem is a finite $D$-dimensional Hilbert space (i.e.
qudit), although each subsystem in superconducting processors is infinite-dimensional.
Even when we use continuous variable codes such as the cat code~\cite{ofekExtendingLifetimeQuantum2016}, there are proposals~\cite{chamberlandBuildingFaultTolerantQuantum2022} to concatenate them with the surface code so that we can suppress errors by increasing the number of subsystems.
Apart from the fact that finite-dimensional codes and quantum algorithms are more mature, there are a few more physical reasons that we consider qudits.
For popular components of superconducting processors such as transmon, fluxonium and resonators, the coherence times of higher energy levels are worse compared to the lower ones.
Therefore, we are motivated to keep the quantum information in the subspace with the best coherence times.
In this work, we will follow the convention and  use the term ``superconducting qubits'' to denote these infinite-dimensional components.
The conflict between the infinite-dimensional components and finite-dimensional computational subspaces creates many interesting design decisions and problems for multi-qubit processors.
We will discuss some of them in the following subsection.

\subsection{Hamiltonian and the computational subspace}
\label{sec:hamiltonian_comp_subspace}

In this work, we consider systems consisting of superconducting qubits.
The total Hilbert space $\mathcal{H}$ is a tensor product defined as
\begin{equation}
\mathcal{H} = \bigotimes_i \mathcal{Q}_i,
\label{eq:product_basis_decomposition}
\end{equation}
where $\mathcal{Q}_i$ represents the Hilbert space of a single superconducting qubit, such as a transmon qubit~\cite{kochChargeinsensitiveQubitDesign2007} or a fluxonium qubit~\cite{manucharyanFluxoniumSingleCooperPair2009}.
We refer to the bases of $\mathcal{H}$ (or its subspace) formed by tensor products of states $\mathcal{Q}_i$ as product bases.
Many of today's superconducting processors can be described by 2-local Hamiltonians of the following form:
\begin{equation}
H(t) = \sum_i H^{(1)}_{i} (t) + \sum_{i,j} H^{(2)}_{ij},
\label{eq:2local_hamiltonian}
\end{equation}
Here, $H^{(1)}_{i}(t)$ and $H^{(2)}_{ij}$ denote the single-qubit and two-qubit Hamiltonians, respectively.
$H^{(1)}_{i} (t)$ acts solely on $\mathcal{Q}_i$, while $H^{(2)}_{ij}$ acts on both $\mathcal{Q}_i$ and $\mathcal{Q}_j$.
Throughout this work, we assume that $H^{(2)}_{ij}$ only influences nearest- and next-nearest-neighbor qubits, although there may exist non-zero couplings between other qubit pairs in reality.
The time independence of the two-qubit interaction Hamiltonians arises from the use of capacitive and inductive couplings between solid-state circuit elements to realize the interaction terms in $H^{(2)}_{ij}$.
The Hamiltonian for a single qubit with time-dependent controls has the following form:
\begin{equation}
H^{(1)}_{i} (t) = H^{(1)}_{i} (0)+\sum_{k=1}^{N_c}f_{i,k}\left(t\right)C_{i,k},
\label{eq:single_qubit_control_hamiltonian}
\end{equation}
Here, $C_{i,k}$ and $f_{i,k}$ represent the control operators and control fields, respectively.
As a convention, we set $f_{i,k}(t) = 0$ for $t\leq 0$, ensuring that $H(0)$ corresponds to the idling Hamiltonian without any applied control.

If we can disable the interactions $H^{(2)}_{ij}$ in~\autoref{eq:2local_hamiltonian}, it is natural to construct a computational subspace $\mathcal{C}_\mathrm{bare}$ by taking the tensor product of $k$ eigenvectors of each $\mathcal{Q}_i$, where $k$ is set to 2 in this study. This specific product basis of $\mathcal{C}_\mathrm{bare}$ will be denoted as $B_\mathrm{bare}$.
With a specific computational basis, we can calculate the matrix representation of operations such as unitary gates and measurements, and subsequently determine the fidelities of these operations.
However, when there are always-on interactions, we often need to consider the system dynamics under a rotation of the Hamiltonian $H\rightarrow U^{\dagger}HU$.
A common choice of rotation is performing a diagonalization, denoted as $U^{\dagger}H(0)U = H_\mathrm{diag}$.
We assume that the eigenvectors of $H(0)$ are close to the product basis states in $B_\mathrm{bare}$, allowing us to unambiguously label each eigenvector as a computational state.
This enables us to obtain the computational subspace $\mathcal{C}_\mathrm{eigen}$ and its basis $B_\mathrm{eigen}$.
When computing the matrix representation of operations and their fidelities under $B_\mathrm{eigen}$, the results will differ from those computed under $B_\mathrm{bare}$. However, if all physical operations themselves remain unchanged, for example, by applying the exact same controls to the qubits and readout resonators, then the aforementioned differences are merely virtual, as the measurement outcomes remain unaffected.
Unfortunately, there is currently no scalable method available to simulate logical error rates in such a way that the result remains constant regardless of the chosen subspace $\mathcal{C}$. The approach utilized in this study necessitates the manual selection of a subspace.
There are several reasons why it may be advantageous to consider a rotated computational subspace, such as $\mathcal{C}_\mathrm{eigen}$:
\begin{enumerate}
    \item Certain operations might exhibit higher fidelities when applied within rotated subspaces. For instance, the study in~\cite{pommereningWhatMeasuredWhen2020} demonstrates that for an experiment-inspired toy model, measurement operations exhibit higher fidelity within an entangling basis, as compared to the product basis $B_\mathrm{bare}$.
    \item
    In general, the interaction terms $H^{(2)}_{ij}$ do not preserve the subspace $\mathcal{C}_\mathrm{bare}$, which can lead to leakage, a troublesome issue. However, by implementing specific rotations, it is possible to achieve a preserved subspace. For example, the subspace $\mathcal{C}_\mathrm{eigen}$ consisting of eigenstates of $H(0)$ remains preserved during idling.
    \item A widely used method for deriving a rotation \(U\) that approximately results a preserved computational subspace \(\mathcal{C}\) is by employing a (perturbative) series, as highlighted in~\cite{bravyiSchriefferWolffTransformation2011,liNonperturbativeAnalyticalDiagonalization2022}. The Hamiltonian, as described in \autoref{eq:2local_hamiltonian}, can be crafted such that several undesirable terms in the effective Hamiltonian series negate each other. A detailed exploration of this is presented in~\autoref{sec:coupling_design} and~\autoref{sec:perturbation_series}.
\end{enumerate}

\subsection{Coupling designs with low 2-body residual interactions}
\label{sec:coupling_design}

For Hamiltonians of the type outlined in \autoref{eq:2local_hamiltonian}, it is likely that the probabilities of weight-$k$ errors diminish as $k$ increases.
This will be discussed further in the subsequent subsction~\autoref{sec:eigenvalue_crosstalk}. Notably, several coupling designs ensure that weight-2 coherent errors during idling operations are minimized or even eradicated. In scenarios where these designs are applied and decoherence errors remain insignificant, it becomes conceivable that weight-$k$ errors, where $k\geq 3$, predominantly contribute to logical errors. This stands in stark contrast to the typical noise models employed in QEC simulations. As previously emphasized, high-weight errors can be considerably more detrimental, underscoring our motivation for conducting comprehensive simulations in this study that account for such errors.

One family of such schemes is known as the tunable couplers~\cite{gellerTunableCouplerSuperconducting2015,mundadaSuppressionQubitCrosstalk2019, yanTunableCouplingScheme2018, xuHighFidelityHighScalabilityTwoQubit2020,stehlikTunableCouplingArchitecture2021, collodoImplementationConditionalPhase2020}.
Here, we will describe some common steps in constructing such couplers.
The first step involves inserting an ancillary coupler qubit, denoted $A_{ij}(x_{ij})$, between each pair of computational qubits, denoted $Q_i$ and $Q_j$.
The coupler qubits are tunable, and $x_{ij}$ is the tunable parameter of $A_{ij}(x_{ij})$.
We then add interactions between $A_{ij}(x_{ij})$ and $Q_{i(j)}$ to the Hamiltonian~\autoref{eq:2local_hamiltonian}.
The coupler qubits $A_{ij}(x_{ij})$ are not used to increase the dimension of the computational subspace $\mathcal{C}$; instead, they are kept in their ground states.
More accurately, $\mathcal{C}$ is approximately spanned by the eigenstates of $H(0)$ that are close to product states where $A_{ij}(x_{ij})$ are in their ground states.
To calculate $\mathcal{C}$ and the corresponding effective Hamiltonians $H_{\mathrm{eff, idle}}$ for the idling Hamiltonian $H(0)$, we can use methods such as the Schrieffer-Wolff (SW) transformation~\cite{bravyiSchriefferWolffTransformation2011}.
We explain this further in~\autoref{sec:perturbation_series}.
By constraining $A_{ij}(x_{ij})$ to their ground states, $H_{\mathrm{eff, idle}}$ appears as a Hamiltonian that only acts on the effective computational qubits ${Q'_i }$, ideally with no interaction between them with proper values of the tunable parameters $\{x_{ij}\}$.
There are also schemes in which the introduced ancillary subsystems are not tunable.
For instance, in~\cite{kandalaDemonstrationHighFidelityCnot2021}, the authors introduced a fixed-frequency bus resonator to reduce the $ZZ$-crosstalk between two transmon qubits.

Introducing tunable ancillary subsystems is not the only approach for engineering effective Hamiltonians in computational subspaces.
In~\cite{nguyenBlueprintHighPerformanceFluxonium2022}, the authors created a two-fluxonium system with both capacitive and inductive coupling that eliminates $ZZ$-crosstalk.
While eventual fabrication uncertainties prevent achieving exactly zero $ZZ$-crosstalk in real devices, the two-fluxonium system with mixed coupling still significantly reduces it, compared to systems with only one type of coupling.
In~\autoref{sec:example_design_control_schemes}, we will demonstrate our workflow by performing numerical simulations on this design.

\subsection{Eigenvalue crosstalk}
\label{sec:eigenvalue_crosstalk}

We can analyze the idling operation of the time-independent Hamiltonian $H(t) = H(0)$ by performing diagonalization and looking at its eigenvalues.
We will assume that the eigenvectors are close to the product basis states and therefore we can unambiguously label the eigenvector corresponding to each state in the product computational basis $B_\mathrm{bare}$.
In general, the diagonalized Hamiltonian restricted to such eigenvectors can be written as
\begin{equation}
    H_\mathrm{diag}=\sum_{\vec{b}} c_{\vec{b}} Z_1^{b_1} Z_2^{b_2}\ldots Z_N^{b_N} \equiv \sum_{\vec{b}} c_{\vec{b}} Z(\vec{b}) .
    \label{eq:zz_crosstalk_hamiltonian_general}
\end{equation}
This is called the Walsh-transform analysis in~\cite{berkeTransmonPlatformQuantum2022}.
Numerical calculation is also given in~\cite{berkeTransmonPlatformQuantum2022} for coupled transmon systems, where it is shown that in general $c_{\vec{b}}\neq 0$.
It is reasonable to assume that we can prepare states which are close to the uniform superposition of eigenvectors (e.g. $\ket{+}^{\otimes n}$) in the computational subspace.
For such states, evolving under $H_\mathrm{diag}$ can generate entangling dynamics.
The unitary operator of idling with $H_\mathrm{diag}$ for a period $t$ is given by:
\begin{equation}
    e^{-i H_\mathrm{diag} t} = \prod_{\vec{b}} e^{-i c_{\vec{b}} Z(\vec{b}) t},
\end{equation}
where we assume $\hbar = 1$.
The terms $e^{-i c_{\vec{b}} Z(\vec{b}) t}$ with $w(\vec{b})=1$ can be compensated by virtual-$Z$~\cite{mckayEfficientGatesQuantum2017,chenPhysRevResearch2023} or real $Z$-rotations.
We define the weight, $w(\vec{b})$, of vector $\vec{b}$ as the count of its non-zero elements.
The terms with $w(\vec{b})=2$ are usually called ``$ZZ$-crosstalk'' in literatures.
In this work, we will use ``eigenvalue crosstalk'' to denote the effects caused by $c_{\vec{b}} Z(\vec{b})$ with $w(\vec{b})\geq 2$.

For general Hamiltonians with weak couplings, $|c_{\vec{b}}|$ decay with $w(\vec{b})$.
Now let us assume that, owing to a good coupling design, $c_{\vec{b}} \approx 0$ for $w(\vec{b}) = 2$, and the dominating terms are $c_{\vec{b}}$ with $w(\vec{b}) = 3$.
If we opt to calculate the Pauli errors in the energy eigenbasis (see~\autoref{sec:convert_unitary_error}), the errors with the highest probabilities are $Z(\vec{b})$ with $w(\vec{b}) = 3$.
We will provide a concrete example with such behavior in~\autoref{sec:results_walsh_transform}.
This scenario is quite different from standard depolarizing noise models such as SD6 and SI1000 (superconducting-inspired 1000 ns cycle)~\cite{gidneyFaultTolerantHoneycombMemory2021} , which are commonly employed in predicting QEC performance through simulation.
While the eigenvalue crosstalk values do not provide direct insight into the error rates of simultaneous operations (see~\autoref{fig:simultaneous_1q_correlated_error} and related discussions), they do indicate the presence of high-weight errors, even in the absence of applied control.

In the following subsection~\autoref{sec:perturbation_series}, we are going to give some intuition about the high-weight errors from the perspective of perturbative expansions.
Specifically, this will shed light on the persistence of high-weight errors even when nearest-neighbor 2-body crosstalks have been cancelled.

\subsection{Perturbative expansions}
\label{sec:perturbation_series}

While methods based on exact time evolution tend to be more precise and perturbative expansions may not always converge as the number of qubits increases, the latter can still offer valuable insights into the systems. In particular, perturbative expansions help us design coupling schemes that reduce weight-2 coherent errors of the idling operation to near or exact zero, as we mentioned in~\autoref{sec:coupling_design}.
Another reason to consider perturbative expansions is that, in general, $H(t)$ does not preserve the product computational subspace $\mathcal{C}_\mathrm{bare}$.
Perturbative expansions can be employed to identify a rotated subspace that approximates the preserved subspace of $H(0)$.

In this section, we will discuss the Schrieffer-Wolff (SW) transformation~\cite{bravyiSchriefferWolffTransformation2011}, although the main idea should be applicable to other perturbative expansions as well.
We treat the 2-body terms in~\autoref{eq:2local_hamiltonian} as a perturbation and rewrite the Hamiltonian at the idling point as
\begin{equation}
    H(0) = \sum_i H^{(1)}_{i} (0)  + \epsilon \sum_{i,j } H^{(2)}_{ij}.
\end{equation}
Then, the expansion series for the effective Hamiltonian can be written as
\begin{equation}
    H_{\mathrm{eff}} = \sum_{k=0}^{\infty}\epsilon^k H_{\mathrm{eff},k}.
    \label{eq:schrieffer_wolff_expansion}
\end{equation}
The expression for $H_{\mathrm{eff},k}$ can be found in~\cite{bravyiSchriefferWolffTransformation2011}.
A noteworthy property is that $H_{\mathrm{eff},n}$ is a sum of operators which are at most $(k+1)$-local.
If we only want to perform single-qubit and two-qubit gates, then these $k$-local operators with $k\geq 3$ are unwanted.
These $k$-local operators can directly lead to weight-$k$ errors, in the sense that the unitary operator of a very short time evolution $\exp (-iH_\mathrm{eff}\Delta t) \approx 1 -iH_\mathrm{eff}\Delta t$ contains $k$-local terms.
We also note that there are some similarities between expansion~\autoref{eq:schrieffer_wolff_expansion} and~\autoref{eq:zz_crosstalk_hamiltonian_general}: Both contain series of $k$-local operators.

An example of applying the SW transformation is the tunable coupler design presented in~\cite{yanTunableCouplingScheme2018}.
The authors employed SW transformation to eliminate the coupler degrees of freedom, thereby obtaining an effective Hamiltonian between computational qubits.
Furthermore, the authors discovered that it is feasible to identify system parameters, allowing the 2-body terms in $H_{\mathrm{eff}}^{(2)} = \sum_{k=0}^{2}\epsilon^k H_{\mathrm{eff},k}$ to be precisely zero by tuning the flux of the coupler qubit.
This cancellation occurs due to the presence of multiple terms in $H_{\mathrm{eff}}^{(2)}$ that negate each other.
It is worth mentioning that non-zero $H^{(2)}_{ij}$ couplings between next-nearest-neighbor qubits contribute to this cancellation.
This intuitive understanding has also been utilized in other coupling designs.
However, it is essential to note that $H_{\mathrm{eff}}$ contains $k$-local operators for $k>2$.
Consequently, the complete cancellation of nearest-neighbor 2-local operators in $H_{\mathrm{eff}}$ does not guarantee the absence of coherent errors.
In fact, the dominant coherent errors may be high-weight correlated errors, necessitating a more cautious approach to their analysis.

We will now briefly examine the applicability of the SW transformation to larger quantum processors.
First, there are two variants of expansion in~\cite{bravyiSchriefferWolffTransformation2011}: a global transformation and a local one.
The convergence of the global transformation is contingent upon the gap between eigenvalues of computational and non-computational states of the unperturbed Hamiltonian.
As more subsystems are added, this gap will eventually disappear, and the norm of the perturbation typically becomes proportional to the total number of subsystems.
Consequently, there is no guarantee that the expansion of the global transformation will converge.
In contrast, the local SW transformation only provides assurance regarding the ground state energy, which is insufficient to infer relevant quantities such as $ZZ$-crosstalks.
Although it is probable that our intuition gained from analyzing small systems will extend to larger ones, a rigorous proof for this belief remains elusive.

\subsection{Simulatable noise model: converting unitary errors}
\label{sec:convert_unitary_error}

There is currently no scalable method to do exact time evolution simulation of the syndrome extraction circuits such as~\autoref{fig:syndrome_circuit}.
As we mentioned, prevalent strategy involves dividing the entire time evolution into operations and converting error of each operation to Pauli operator.
In this subsection, we will discuss how to convert unitary errors to stochastic Pauli errors, and the potential problems of this conversion.
We consider stochastic Pauli errors because we rely on the Gottesman-Knill theorem~\cite{gottesman1998heisenberg} to simulate syndrome extraction circuits for QEC codes with non-trivial sizes.
We can justify the stochastic Pauli error model by potentially using randomized compiling~\cite{wallmanNoiseTailoringScalable2016} to convert general errors into Pauli errors, although it is often not performed in experiments~\cite{acharyaSuppressingQuantumErrors2023,chenCalibratedDecodersExperimental2022,krinnerRealizingRepeatedQuantum2022}, possibly due to extra errors caused by the additional single-qubit gates.
Another reason for not performing randomized compiling is that unitary errors from different rounds of gates can potentially cancel coherently.
For example, in \cite{acharyaSuppressingQuantumErrors2023}, an extra round of simultaneous gates is inserted into the syndrome extraction circuit with the purpose of cancelling unitary errors.

The first step for computing unitary matrices of gates is choosing a computational subspace $\mathcal{C}$ and a basis $B$ of $\mathcal{C}$.
Then, for each state $\ket{\psi_{i}}$ in $B$, we can compute the corresponding final state $\ket{\psi_{\mathrm{fin},i}}$.
It is likely that some final states $\ket{\psi_{\mathrm{fin},i}}$ are not fully in $\mathcal{C}$.
This is called the leakage errors.
Since the focus of this work is not leakage, we will assume that the final states $\ket{\psi_{\mathrm{fin},i}} \in \mathcal{C}$ (see~\cite{acharyaSuppressingQuantumErrors2023} for a stochastic error model with leakage).
In this case, the time evolution can be described by a unitary operator $U_\mathrm{sim}$.
Let us call the target operation matrix $U_{\mathrm{target}}$.
Then, the unitary error is $U_\mathrm{error} = U_{\mathrm{sim}}^{\dag} U_{\mathrm{target}}$.
We can expand $U_\mathrm{error}$ in the Pauli basis $U_\mathrm{error} = \sum_{\vec{j}} a(\vec{j})P_{\vec{j}}$,
where $P_{\vec{j}}$ is a Pauli operator
\begin{equation}
    P_{\vec{j}} = \sigma_{j_1}\otimes \cdots \otimes \sigma_{j_n}, \quad \sigma_{j}=I,X,Y,Z.
 \label{eq:pauli_operator_multiqubit}
\end{equation}
 After applying Pauli twirling~\cite{dankertExactApproximateUnitary2009}, the probability of having a Pauli error is
\begin{equation}
    \mathrm{Prob}(E =P_{\vec{j}} ) = |a(\vec{j})|^2.
    \label{eq:error_prob_from_twirling}
\end{equation}
Since the syndrome measurement circuit requires multiple different gates, we will use $U_{\mathrm{error},i}$, $U_{\mathrm{sim},i}$, and $U_{\mathrm{target},i}$ to represent the corresponding matrices for the gate with index $i$.

However, it is easy to notice that there is freedom in choosing the computational subspace $\mathcal{C}$ and a basis $B$ of $\mathcal{C}$.
Let us first assume that $\mathcal{C}$ is fixed.
Then, when we do a change of basis represented by matrix $V$, the matrices $\{ U_{\mathrm{sim},i} \}$ of a gate set will undergo the transformation
\begin{equation}
    \{ U_{\mathrm{sim},i} \} \rightarrow \{ V^\dagger U_{\mathrm{sim},i} V \}.
    \label{eq:basis_change_gateset}
\end{equation}
Therefore, the choice of basis $B$ will affect the errors $U_{\mathrm{error},i} = U_{\mathrm{sim},i}^{\dag} U_{\mathrm{target},i}$.
By changing $V$, we can readily introduce arbitrarily large errors that may be correlated across multiple qubits.
However, a more intriguing scenario arises when different $V_i$ are needed to minimize the errors of different gates $U_{\mathrm{sim},i}$.
In this case, it is unclear how to select $V$ so that the simulation closest to the correct outputs.
This illustrates an instance of the gauge freedom concept discussed in~\cite{nielsenGateSetTomography2021}.
At the core, the simulation done in the work is gauge-variant, i.e.
the logical error rates depend on the choice of $\mathcal{C}$ and $B$.
Currently, the only gauge-invariant method of simulating QEC circuits is a brute force simulation of the whole circuits, which is not computational feasible beyond very small code distances.
In this work, we will choose $\mathcal{C}$ and $B$ manually.
It is an interesting question whether better methods exist, such as the gauge optimization mentioned in~\cite{nielsenGateSetTomography2021}, or even finding some ways to compute quantities such as logical error rates in a gauge-invariant manner.

\begin{assumption}
    \label{assumption:eigenbasis}
For computing $U_{\mathrm{sim},i}$ and $U_{\mathrm{error},i}$, we will set the computational subspace $\mathcal{C}_\mathrm{eigen}$ based on the eigenbasis of the system's total Hamiltonian at idling point $H(0)$ from~\autoref{eq:2local_hamiltonian} as the basis for computing error rates.
More concretely, we will use the eigenstates which are close to the product subspace $\mathcal{C}_\mathrm{bare}$ introduced in~\autoref{sec:hamiltonian_comp_subspace} to span $\mathcal{C}_\mathrm{eigen}$.
The basis $B_\mathrm{eigen}$ of $\mathcal{C}_\mathrm{eigen}$ will also be chosen in a way so that it is close to the product basis $B_\mathrm{bare}$.
The details will be provided below.
\end{assumption}
For the example considered in this work, the states in the eigenbasis are close enough to the product states in $B_\mathrm{bare}$, and we only consider small amounts of qubits in numerical simulation.
Therefore, we can unambiguously pick the eigenstates $\ket{\psi_i}$ to span $\mathcal{C}_\mathrm{eigen}$ as long as $|\langle \varphi \ket{\psi_i}|>0.9$ for some product state $\ket{\varphi}\in \mathcal{C}_{\mathrm{prod}}$.
Now we still need to choose suitable phases for these eigenstates to form a basis $B_\mathrm{eigen}$ close to $B_\mathrm{bare}$.
Let us use unitary matrix $U_B$ to describe basis $B$, where each column of $U_B$ corresponds to a vector in $B$.
The action of multiplying different phases to the eigenstates can then be described by the transformation $U_B \rightarrow U_{B} D$.
This will transform the matrices of gates according to~\autoref{eq:basis_change_gateset}.
Since $D$ is a global diagonal matrix, the above transformation can be drastic.

We will use the following procedure to construct $B_\mathrm{eigen}$ which is close to $B_\mathrm{bare}$.
We can first write the idling Hamiltonian matrix $H(0)$ in the product basis.
By diagonalizing $H(0)$, we can obtain the eigenstates and thus the matrix $U_{B}$ in the product basis, where $B$ is a basis of $\mathcal{C}_\mathrm{eigen}$ consisting eigenstates of $H(0)$.
To make the distance between $U_{B_\mathrm{eigen}} = U_{B}D$ and $U_{B_\mathrm{bare}}=I$ small, we will simply choose $D$ so that the diagonal elements $u_{jj} = r_j e^{i\theta_j}$ of $U_{B} D$ satisfy $e^{i\theta_j}=1$.

\subsection{Simulatable noise model: correlated errors}
\label{sec:noise_model}

\begin{figure}
    \includegraphics[width=0.8\linewidth]{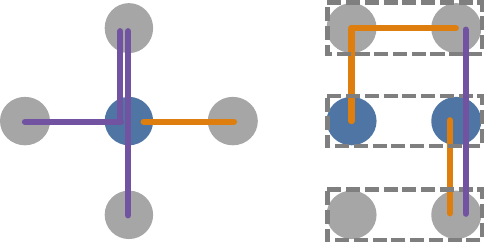}
    \caption{The noise models we consider for single- and two-qubit gates.
In the figures, each circle represents a qubit, and each line represents a correlated Pauli error affecting the qubits it passes through.
We illustrate errors in LCPEM corresponding to the gate locations with the blue qubits.
In the left figure, the orange and purple lines represent Pauli errors affecting two and three simultaneous single-qubit gates, respectively.
In the right figure, dashed boxes represent the locations of 2-qubit gates.
The orange and purple lines represent Pauli errors affecting qubits belonging to two and three pairs of simultaneous two-qubit gates, respectively.
Using the notation defined in~\autoref{sec:noise_model}, we will call the errors corresponding to purple and orange lines as $2(3)$-location errors correspondingly.
Under the uniform LCPEM error model specified in Assumption~\ref{assumption:uniform_lcpem}, equal probabilities will be assigned to the Pauli errors represented by the lines with the same color in each figure.
  \label{fig:local_correlated_noise_model}}
\end{figure}

In this subsection, we describe our approach to handling correlated errors in noise models and the approximations we employ.
The first approximation involves performing time evolution simulations on a fixed small number of qubits and utilizing the resulting information to construct the noise model for the whole lattice.
A specific example of how we choose a subset of qubits is provided in~\autoref{fig:frequency_pattern_and_simul_cnot}.
This approximation is considered unsafe since we cannot guarantee that a simulation with a greater number of qubits will yield only negligible differences.

Additionally, we need a family of noise models that enables extrapolation of errors from a few simulated qubits to a varying number of qubits in surface codes.
In this study, we consider noise models akin to the ballistic and diffusive models outlined in~\cite{nickersonAnalysingCorrelatedNoise2019}.
We refer to our models as the locally correlated Pauli error models (LCPEM), which is defined by the following process of sampling errors.
For each gate, we independently determine whether it induces errors.
If errors arise, they can affect qubits within the gate's neighborhood.
This key distinction separates LCPEM from the standard depolarizing noise models such as SD6, where Pauli errors can only occur on the qubits to which the gate is applied.

To assign probabilities to correlated errors that affect different numbers of qubits, let us first review the concept of locations in fault tolerance (e.g., see~\cite{gottesmanIntroductionQuantumError2010}).
A location represents an instantiation of one of the operations (e.g., state preparation, gates, measurements, idling) in the circuits.
This concept is valuable because fault-tolerant circuits are typically designed to withstand errors in up to $m$ locations.
For instance, a well-designed syndrome extraction circuit for surface codes (e.g.,~\autoref{fig:syndrome_circuit}) can tolerate errors in fewer than $d/2$ locations~\cite{yoderSurfaceCodeTwist2017}.
This might imply that the number of faulty locations is more critical than the number of faulty qubits.
Further studies are needed to determine which factor is more influential to the logical error rates.
Additionally, we note that we will only consider space-like correlated errors, meaning that we will not have correlated errors on locations from different time steps in the syndrome measurement circuits.
The error rate of a gate is also assumed to be time-independent.
We will call a correlated Pauli error on $k$ locations a $k$-location error.

In~\autoref{fig:local_correlated_noise_model}, we illustrate the neighborhoods surrounding single- and two-qubit gates.
We assign a 5(6)-qubit Pauli error distribution, $q_i(\vec{j})$, to the central 1(2)-qubit gate labeled $i$.
For each gate $i$ in the syndrome extraction circuit~\autoref{fig:syndrome_circuit}, $\vec{j}_i$ is sampled from corresponding distribution.
This represents a multi-qubit error $P_{\vec{j}_i}$ with the form described in~\autoref{eq:pauli_operator_multiqubit} happened on a neighborhood of gate $i$.
As long as $P_{\vec{j}_i}\neq I$, we require the restriction of $P_{\vec{j}_i}$ to the central gate location is also not $I$.
This requirement ensures that we will not count errors repeatedly.
We choose to limit ourselves to $P_{\vec{j}_i}$ that have support on at most three locations inside the corresponding neighborhood shown in~\autoref{fig:local_correlated_noise_model}.
After completing the sampling process, we compute the product of all Pauli errors for one round of simultaneous gate operations to obtain the total error for that round, given by
\begin{equation}
E=\prod_i P_{\vec{j}_i}.
\label{eq:multiplying_simul_gate_error}
\end{equation}
Since changing the multiplication order only introduces a possible $-1$ multiplicative coefficient to $E$, it does not affect the QEC simulation.
Consequently, the multiplication can be performed in any order.
It is worth noting that the errors corresponding to a gate in the sampling procedure are not necessarily the same as the errors when the gate is applied in isolation.
In fact, it is quite likely that the errors for isolated gates and simultaneous gates differ.

We will further restrict ourselves to a subclass of LCPEM, which we refer to as uniform LCPEM, since LCPEM still contains too many parameters for our purpose.
For uniform LCPEM, we assume that errors affecting the same number of neighboring locations have equal probabilities.
In summary, we consider the following uniform LCPEM:
\begin{assumption}
    For each gate involved in simultaneous single-qubit operations, we assume a probability $p^{(1)}_k$ of having random Pauli errors on each combination of $k$ neighboring gates.
Any such $k$-location Pauli error is presumed to have equal probability, and they sum up to $p^{(1)}_k$.
Similarly, for simultaneous two-qubit operations, we assume the probability of having errors on $k$ neighboring gate locations is $p^{(2)}_k$.
We again assign a uniform error probability to each $k$.
We assume that simultaneous single and two-qubit operations in different time steps of~\autoref{fig:syndrome_circuit} have the same parameters $p^{(j)}_k$.
We refer to the error events corresponding to $p^{(j)}_k$ as $k$-location errors.
For simultaneous state preparation and measurement operations, we only consider single-qubit independent errors and assume their error rates are all equal to the same parameter $p_\mathrm{spam}$.
    \label{assumption:uniform_lcpem}
\end{assumption}

To provide clarity on the definition of uniform LCPEM, we present examples in~\autoref{fig:local_correlated_noise_model}.
For instance, in the left panel, the probabilities of all 3-location Pauli errors corresponding to the two purple lines are equal.
There are in total 6 different configurations of such 3-location errors which pass through the central qubit, and all corresponding Pauli error probabilities sum to $p^{(1)}_3$.
In the right panel, the probabilities of all 2-location Pauli errors corresponding to the two orange lines are equal.

We want to mention that state preparation and measurement operations are also likely to have some high-weight correlated errors.
For example, different aspects of correlated errors in measurement have been discussed in~\cite{bravyiMitigatingMeasurementErrors2021,pommereningWhatMeasuredWhen2020,lienhard2022deep}.
Accurate simulation of these operations is beyond the scope of this work.

\subsection{Estimating the effects of high-weight errors in the noise model}
\label{sec:effects_high_weight_errors}

In this subsection, we will provide some estimates for the effects of high-weight errors.
First, we can argue that error events with probabilities much smaller than the logical error rates have negligible effects.
More precisely, let us assume that in a simulation of a surface code memory experiment, the expected number of LCPEM weight-$k$ errors is $\bar{n}_k$, and the expected number of logical errors is $\bar{n}_{\mathrm{L}}$.
If $\bar{n}_k \ll \bar{n}_{\mathrm{L}}$, then it is evident that the majority of logical errors are not related to those weight-$k$ errors.
Therefore, for those low probability errors to have any effects, we need to reduce the logical error rates to similar magnitudes by reducing decoherence errors or increasing the code distance when under the threshold.

Next, we investigate a scenario in which only high-weight errors are present in LCPEM, while decoherence errors are absent.
We analyze a toy model by considering single round QEC with LCPEM and perfect syndrome extraction.
The LCPEM we consider aligns with the ballistic noise model described in~\cite{nickersonAnalysingCorrelatedNoise2019}.
Specifically, for each qubit on the vertical link, there is a probability $p$ that $X$ errors occur on it and the qubit to its right.
Similarly, for each qubit on the horizontal link, there is a probability $p$ that $X$ errors occur on it and the qubit on its bottom.
We choose this correlated noise model because it allows us to obtain analytical results.
Additionally, we consider surface codes with periodic boundary conditions, i.e., toric codes, where the code distance $d$ is assumed to be even.

\begin{figure}
    \includegraphics[width=0.9\linewidth]{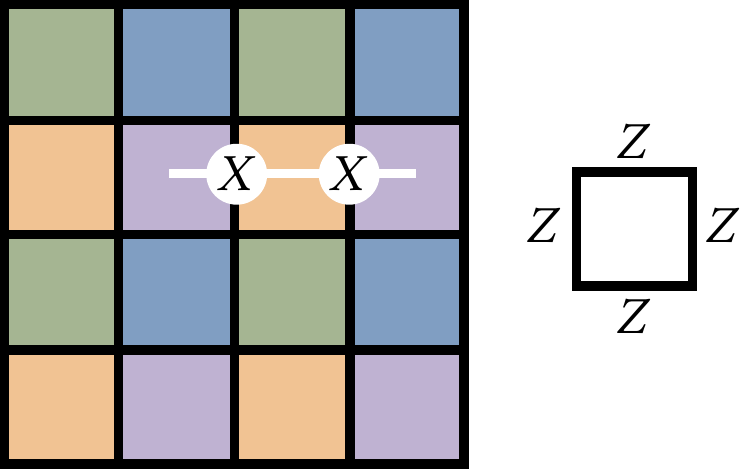}
    \caption{In the figure, each edge is a qubit and each plaquatte corresponds to a $Z$-type stabilizer check.
For the error model which only has length-2 ballistic $X$ error, we can do decoding for plaquettes with different color separately.
This is because each error event will always lead to a pair of $Z$-type stabilizer detection event in the neighboring plaquettes with the same color.
\label{fig:symmetry_decoding}}
\end{figure}
First, let us discuss the threshold for the above correlated error model.
By using the argument in~\cite{brown2023conservation}, we can map the decoding problem of the above correlated noise model to 4 standard decoding problems of surface code with independent bit-flip errors.
Therefore, the threshold of the correlated noise model is identical to the threshold of surface code, which is $p_{\mathrm{MWPM}}\approx 0.103$~\cite{wangConfinementHiggsTransitionDisordered2003} if we use the matching decoder.
This is somewhat surprising since not only the errors are correlated, the marginalized physical error rate for each qubit is also roughly doubled.
We will summarize the argument for completeness.
For length-2 $X$ error strings described above, we can see in~\autoref{fig:symmetry_decoding} that the strings can only start and end in the plaquettes with same colors.
Therefore, we can do decoding for plaquettes with each color separately.
This behavior can be viewed from the perspective of symmetry~\cite{brown2023conservation} by noticing that length-2 $X$ error strings commutes with the product of the $Z$-stabilizer plaquette checks corresponding to each color.

For the low error rate regime~\cite{watsonLogicalErrorRate2014,fowlerSurfaceCodesPractical2012}, let us first consider a distance $d$ surface code with independent weight-1 bit-flip errors, where we assume that the distance $d$ of the code is fixed and the error rates $p\rightarrow 0$.
Then, the logical error rate has the form
\begin{equation}
    P_{\mathrm{logical}, d} (p) = \frac{2d}{2} \binom{d}{d/2} p^{d/2} + O(p^{d/2+1}),
\end{equation}
where $\binom{d}{d/2}$ is the number of different configurations in which $d/2$ errors occur along a shortest length-$d$ logical operator.
The factor of $2d$ in the numerator of $\sfrac{2d}{2}$ corresponds to the total number of such shortest logical operators.
The factor of 2 in the denominator arises because there are two error configurations on a line with the same syndrome and the same number of errors $d/2$.
As a result, there is a 50\% chance of correct decoding.

Now let us switch back to the weight-2 ballistic noise model.
As previously discussed, when considering the weight-2 ballistic noise model, the decoding problem can be divided into 4 identical copies.
Each copy exhibits a logical error rate of $P_{\mathrm{logical}, d/2}$.
When it is very small, the total logical error rate of 4 decoding problems is approximately $4P_{\mathrm{logical}, d/2}$.
So we can conclude that the main effect of weight-2 errors is a reduction in the distance $d$ by half, which aligns with our intuition.

For the above error model, we can observe that high-weight errors have large detrimental effects in the low-error regime but no effect on the value of the threshold.
We expect this behavior to carry over to general LCPEM qualitatively.
Firstly, high-weight errors in LCPEM will reduce the effective code distance and, consequently, impact the logical error rates in the low-error regime.
On the other hand, the value of the threshold is a property of the limit as $d\rightarrow \infty$.
Therefore, reducing the effective distance from $d'=d$ to $d'=d/k$, does not necessarily change the threshold.
Based on the above discussion, it is likely that optimizing the processor to reach the threshold and optimizing for low logical error rates are two separate optimization problems.
When we aim only to reach the threshold, we may be more lenient about high-weight errors and thus have an incentive to increase the couplings $H^{(2)}_{ij}$ to have faster gates and lower decoherence errors.
In this work, we will not attempt to numerically evaluate or optimize the QEC performance in the regime where we are close to the threshold and $H^{(2)}_{ij}$ are much larger compared to typical values.
This is because we are unsure about the numerical accuracies in the above regime, and designing decoders for general LCPEM noise models is beyond the scope of this work.
We note that the choice of decoders affects both the threshold and the logical error rates.

To achieve fault-tolerant quantum computation, the processors need to operate under the threshold with a comfortable margin.
Therefore, the weights of errors can play a large role in determining the logical error rates.
On the other hand, as we will shown in~\aref{sec:fidelity_high_weight_errors}, the commonly used measure average fidelity is not affected by the weights of errors.
This implies that we cannot rely solely on average fidelity to predict the performance of error correction unless we are confident that the high-weight errors are negligible.
For this reason, we will simulate the syndrome extraction circuits with high-weight errors to obtain a better estimation of logical error rates.

\subsection{Discussion of previous works}

To the best of our knowledge, there is no well-stated condition and proof of a fault-tolerance threshold theorem for typical planar superconducting processors with a Hamiltonian of the form~\autoref{eq:2local_hamiltonian}. There are two significant challenges that have yet to be addressed.
Firstly, we do not have a proof for threshold theorem when we replace stochastic errors by coherent errors in the circuit-level noise models such as SD6, 
but we should also note that several studies~\cite{bravyiCorrectingCoherentErrors2018,venn2023coherent, marton2023coherent} suggesting the persistence of error thresholds under various simplified scenarios.
Even if we assume that these coherent errors can be fully converted into stochastic errors through randomized compiling~\cite{wallmanNoiseTailoringScalable2016}, we still do not know how to bound the decay speed of correlated errors resulting from the time evolution of the Hamiltonian in~\autoref{eq:2local_hamiltonian}.
The unknown decay speeds will also be an issue in the threshold theorems proven for concatenated codes~\cite{aharonovFaultTolerantQuantumComputation2008,aharonovFaultTolerantQuantumComputation2006}.

For the circuit-level noise models, the most well-studied setting is that each operation creates an independent stochastic error event on the qubits which the operation acts on~\cite{fowlerSurfaceCodesPractical2012,dennisTopologicalQuantumMemory2002,gidneyFaultTolerantHoneycombMemory2021}.
For brevity, we will refer to these models as the circuit-level depolarizing noise models.
As mentioned earlier, this work focuses on noise models that exhibit higher spatial correlation.
For instance, individual stochastic error events may result in errors on qubits that were not targeted by the operation.
Learning of such noise models are considered in~\cite{harperEfficientLearningQuantum2020,harperLearningCorrelatedNoise2023}.
In particular, \cite{harperLearningCorrelatedNoise2023} also mention the importance of considering correlated noise for predicting the sub-threshold behavior.
There are also previous works about analyzing and decoding non-local correlated noises~\cite{chubbStatisticalMechanicalModels2021,nickersonAnalysingCorrelatedNoise2019, maskaraAdvantagesVersatileNeuralnetwork2019, huangEfficientParallelizationTensor2021}.

Several studies have investigated QEC simulations on superconducting processors, including ~\cite{acharyaSuppressingQuantumErrors2023,chenExponentialSuppressionBit2021, krinnerRealizingRepeatedQuantum2022,obrienDensitymatrixSimulationSmall2017}.
Of particular note, Google Quantum AI~\cite{acharyaSuppressingQuantumErrors2023} examine spatially correlated errors that go beyond circuit-level depolarizing noise models such as SD6.
Our contribution differs from previous works in two key ways.
First, we provide a general argument for the presence of high-weight correlated errors on superconducting processors, as opposed to only focusing on the specific system analyzed in~\cite{acharyaSuppressingQuantumErrors2023}.
Additionally, we offer a theoretical explanation for some of the effects of high-weight errors, complementing the numerical simulations found in our works.
Second, we outline a route to do gradient optimization which takes into account the effects of high-weight errors.
This presents a promising avenue for mitigating the effects of these errors in future QEC simulations on superconducting processors.

There are several recent works~\cite{leungSpeedupQuantumOptimal2017, shillitoFastDifferentiableSimulation2021,puzzuoli2023algorithms} on fast differentiable simulation of time-evolution.
In particular, the method presented in~\cite{puzzuoli2023algorithms} is closely related to time-dependent perturbation theory, which may make it a suitable method for computing the amplitudes of correlated errors.
However, we do not provide a detailed comparison of the application of these methods to the simulation of multi-qubit systems in this work as it is beyond the scope of our study.

\section{Workflow for computing logical error rates from Hamiltonians}
\label{sec:workflow}

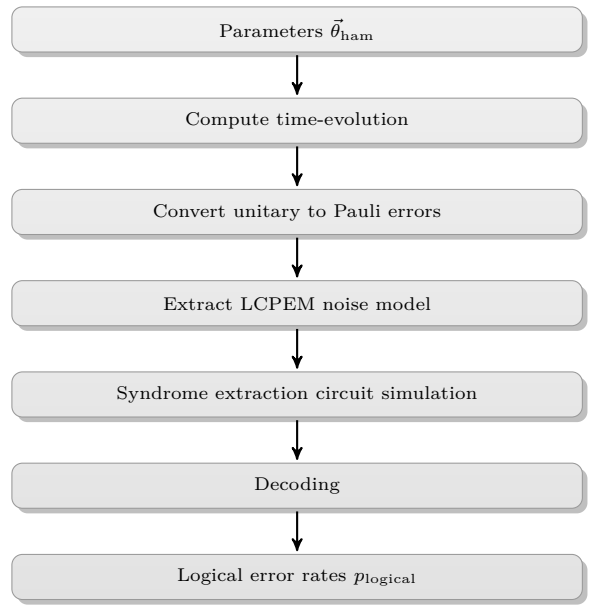
\begin{figure}
    \centering
\begin{tikzpicture}[
  node distance=0.6cm and 0.6cm,
  box/.style={draw, rectangle, rounded corners, minimum width=0.8\linewidth, minimum height=0.6cm, text width=0.8\linewidth, align=center, drop shadow, draw=gray!80, font=\scriptsize, middle color=white},
  arrow/.style={thick,->,>=stealth', shorten >=1pt, shorten <=1pt}
]
\definecolor{colorStart}{RGB}{240,240,240}
\definecolor{colorEnd}{RGB}{220,220,220}

\node (input) [box, top color=colorStart!100!colorEnd, bottom color=colorStart!90!colorEnd] {Parameters $\hamparam$};
\node (step1) [box, below=of input, top color=colorStart!90!colorEnd, bottom color=colorStart!80!colorEnd] {Compute time-evolution};
\node (step2) [box, below=of step1, top color=colorStart!80!colorEnd, bottom color=colorStart!70!colorEnd] {Convert unitary to Pauli errors};
\node (step3) [box, below=of step2, top color=colorStart!70!colorEnd, bottom color=colorStart!60!colorEnd] {Extract LCPEM noise model};
\node (step4) [box, below=of step3, top color=colorStart!60!colorEnd, bottom color=colorStart!50!colorEnd] {Syndrome extraction circuit simulation};
\node (step5) [box, below=of step4, top color=colorStart!50!colorEnd, bottom color=colorStart!40!colorEnd] {Decoding};
\node (output) [box, below=of step5, top color=colorStart!40!colorEnd, bottom color=colorStart!30!colorEnd] {Logical error rates $\ler$};

\draw[arrow] (input) -- (step1);
\draw[arrow] (step1) -- (step2);
\draw[arrow] (step2) -- (step3);
\draw[arrow] (step3) -- (step4);
\draw[arrow] (step4) -- (step5);
\draw[arrow] (step5) -- (output);

\end{tikzpicture}
\caption{This flowchart illustrates the process for forward computation of the logical error rates from the parameters $\hamparam$. Gradients are computed using backpropagation for the first four steps and estimated using finite differences for the last two.}
\label{fig:workflow}
\end{figure}

By grouping terms in~\autoref{eq:2local_hamiltonian} and~\autoref{eq:single_qubit_control_hamiltonian}, we obtain the following parameterized Hamiltonian:
\begin{equation}
    H\left(\vec{h}, \vec{c},t\right)=
    H_{\mathrm{idle}}\left(\vec{h}\right)+\sum_{k=1}^{N_c}f_{k}\left(\vec{c},t\right)C_{k}\left(\vec{h}\right),
    \label{eq_hamiltonian_parameter}
\end{equation}
where  $\vec{h}$ and $\vec{c}$ represent the device and control parameters, respectively.
We represent these variables as vectors to emphasize that they comprise multiple parameters.
$H_{\mathrm{idle}}\left(\vec{h}\right) = H\left(\vec{h}, \vec{c},0 \right)$ is the Hamiltonian when the processor is idling, i.e., when no controls are applied.
The combined set of parameters ${\vec{h},\vec{c}}$ will henceforth be referred to as Hamiltonian parameters $\hamparam$.
In this section, we present a workflow for estimate the logical error rates $\ler$ from $\hamparam$, keeping in mind that our estimates are subject to various approximation errors.
Our approach is outlined in~\autoref{fig:workflow}.
We will also discuss how to compute the gradient $\nabla \ler (\hamparam)$ for use in optimization routines.

\subsection{Time-evolution simulation}
The first step is to calculate the time evolution $U_\mathrm{sim}$ of simultaneous 1-qubit and 2-qubit gates using standard numerical methods such as differential equation solvers or Trotter expansions.
In this work, we simulate a small system and thus can employ these straightforward methods.
Also, the simulation is done similarly as~\cite{niIntegratingQuantumProcessor2022} so that we can compute the derivative of $U_\mathrm{sim}$ with respect to $\hamparam$.
This is important not only for minimizing $\ler$, but also for preliminarily optimizing control parameters $\vec{c}$, which ensures that $U_\mathrm{sim}$ approximates the desired gate operation.
We require this preliminary optimization because otherwise we do not know how to assign parameters such as drive frequencies and amplitudes.
A natural approach is to perform an optimization that maximizes the average fidelity between $U_\mathrm{sim}$ and the target operation.
The details of our time evolution simulations are presented in~\aref{sec:detail_time-evolution simulation}.

\subsection{Extract LCPEM noise model parameters}

The step of converting unitary errors to Pauli errors is described in~\autoref{sec:convert_unitary_error}.
After obtaining the Pauli error rates of simultaneous gate operations, we use these values to construct a uniform LCPEM for simulating QEC circuits on larger lattices.
To obtain parameters of uniform LCPEM, we divide the errors $E$ in~\autoref{eq:error_prob_from_twirling} into $k$-location errors with different $k$, and then average over $k$-location errors probabilities $|a(\vec{i})|^2$ to obtain $p^{(j)}_k$ described in Assumption~\ref{assumption:uniform_lcpem}.

To incorporate error probabilities from decoherence into $p_k^{(j)}$, we assume that decoherence can be described by a single small rate $r$.
For each single-qubit and two-qubit gate operations, we set the decoherence error probabilities to $rt_{\mathrm{1q}}$ and $2rt_{\mathrm{2q}}$, respectively.
Thus, for gate operations we have
\begin{equation}
p_{1,\mathrm{total}}^{(j)} = p_{1,\mathrm{unitary}}^{(j)} + jrt_{j\mathrm{q}}.
\label{eq:add_decoherence_error_to_unitary}
\end{equation}
Here, $j=1,2$ represents whether the gate is single-qubit or two-qubit, and $t_{j\mathrm{q}}$ denotes the respective gate time.
The error rate $p_{1,\mathrm{unitary}}^{(j)}$ is obtained by performing twirling on the unitary error from the previous step.

For qubits that are idling in the syndrome measurement circuit (see ~\autoref{fig:syndrome_circuit}), the idling durations are determined based on the operations in the same time step. Specifically, we set the idling times as follows: $t_{\mathrm{1q}}=40$ ns, $t_{\mathrm{2q}}=130$ ns, $t_{\mathrm{reset}}=160$ ns, and $t_{\mathrm{measure}}=500$ ns. The gate times are approximated to be equal to the values used in the time-evolution simulation (refer to~\autoref{table:parameters_before_simple_optimization}). The durations for reset and measurement are determined based on the values reported in ~\cite{acharyaSuppressingQuantumErrors2023}.
We also assume that their error rates are constant, with $p_{\mathrm{reset}}=0.005$ and $p_{\mathrm{measure}}=0.01$.
We acknowledge that these estimations provide a rough approximation for the decoherence and SPAM errors. To achieve a higher level of accuracy, a more meticulous modeling of these errors would be necessary. However, such detailed modeling is beyond the scope of this study.

\subsection{QEC simulation}
\label{sec:qec_sim_workflow}

After obtaining the uniform LCPEM, simulating the surface code syndrome extraction circuits becomes a standard procedure due to the Gottesman-Knill theorem~\cite{gottesman1998heisenberg}.
To describe the procedure in more detail, let us begin with the standard circuit-level depolarizing noise model (e.g.,~\cite{fowlerSurfaceCodesPractical2012}).
For each faulty operation in the circuit, we sample a possible Pauli operator and append it before or after the operation.
The sampled Pauli operator has support only on the qubits corresponding to the faulty operation.
Then, for each round of simultaneous operations, we multiply the sampled Pauli operators together to obtain the total error for the simultaneous operations.
To simulate with the LCPEM, the only difference is that when we sample a Pauli operator for each faulty operation, the operator can act on more qubits than the operation itself.
Afterwards, we can multiply these errors together as described in~\autoref{eq:multiplying_simul_gate_error}.

One debatable choice we make is to assume that the idling operations in the syndrome measurement circuit shown in~\autoref{fig:syndrome_circuit} do not cause correlated errors, i.e., we set $p_{2,3}^{(1)} = 0$ for these operations.
Additionally, when we sample high-weight Pauli operator near the boundaries of the lattice, there is a chance that the support of the operator extends beyond the lattice.
In such cases, we truncate the Pauli operator to act only on the lattice.

For decoding, we use a union-find decoder~\cite{delfosseAlmostlinearTimeDecoding2017}.
We modify the decoder graphs to correspond to the circuit-level error model of the surface code.
More concretely, the decoder graphs are formed by connecting all space-time sites that can be jointly excited by a single circuit fault, in the same way as~\cite{huangFaulttolerantWeightedUnionfind2020}.
We will assign equal weight to all edges in the decoder graph, partially because it is not clear how to optimally assign weights for the LCPEM.

In this work, we will only simulate the quantum memory task in which we begin with $\ket{0}^{\otimes N}$ and conclude by measuring in the $Z$-basis. Therefore, the intended initial logical state is $\bar{\ket{0}}$, and we solely examine whether a logical-$X$ error occurs after decoding.
For the specific syndrome measurement circuit depicted in~\autoref{fig:syndrome_circuit}, since Hadamard gates only apply to ancilla qubits for X-type checks, local errors on these gates do not affect the logical error rate for Z-basis memory experiment.
Consequently, in our numerical simulation, the derivative $\sfrac{\partial \ler}{\partial p_1^{(1)}} = 0$.

\subsection{Gradient computation}

We can estimate the gradients $\nabla \ler (\hamparam)$ as well.
The computation is divided into two parts.
First, for the computation from $\hamparam$ to uniform LCPEM probabilities $p_k^{(j)}$, we can compute the gradients using efficient backpropagation.
This can be done based on~\cite{niIntegratingQuantumProcessor2022}, since the additional steps of converting to Pauli errors and extracting LCPEM probabilities are differentiable.
The cost for this part of the gradient computation is bounded by the cost of computing $p_k^{(j)}$ times a constant $c\approx 5$, making it efficient.

The second part is the computation from ${p_k^{(j)}}$ to $\ler$.
For this, we will estimate the gradients using finite-difference.
Since ${p_k^{(j)}}$ consists of only six real numbers, the overhead caused by doing finite-difference for each of them is not substantial.
We can then combine these two parts using the chain rule.

One non-trivial issue we need to handle in the gradient computation is that some parameters need to be shared, i.e., those parameters always have the same value during optimization.
For example, one way to ensure that some qubit frequency arrangements can be scaled to large 2D processors is to enforce frequency patterns in the layout~\cite{hertzbergLaserannealingJosephsonJunctions2021, dipaoloExtensibleCircuitQEDArchitecture2022}. To address this problem, we will optimize the qubit parameters in the frequency patterns instead of treating the parameters of each qubit as independent variables.
We will discuss how to implement this in software in future work.

\section{An application to fluxonium processors with multi-path coupling}
\label{sec:example_design_control_schemes}

\subsection{Overview of the system and gate schemes}

To demonstrate the above workflow with an example, we consider the fluxonium processor design described in~\cite{nguyenBlueprintHighPerformanceFluxonium2022}.
In their paper, the authors proposed a multipath coupling scheme for fluxoniums which can cancel the two-body $ZZ$ energy eigenvalue crosstalk by having a suitable combination of capacitive and inductive couplings.
More concretely, the time-independent part of processor's Hamiltonian is
\begin{equation}
    H(0) = \sum_i H_{\mathrm{f},i}  + \sum_{\langle i,j \rangle } H_{ij}.
    \label{eq:two_local_fluxonium_Hamiltonian}
\end{equation}
In the above equation, $H_{\mathrm{f},i}$ is the Hamiltonian for individual fluxonium
\begin{equation}
    H_{\mathrm{f},i} = 4E_{\mathrm{C},i} \opn_i^2 + \frac{1}{2} E_{\mathrm{L},i} (\opv_i +\varphi_{\mathrm{ext},i})^2
    -E_{\mathrm{J},i}\cos \left( \opv_i \right),
    \label{eq:hamiltonian_single_fluxonium}
\end{equation}
where $E_{\mathrm{C},i}$, $E_{\mathrm{J},i}$, and $E_{\mathrm{L},i}$ are the charging energy, the Josephson energy, and the inductive energy, respectively.
$\opv_i$ is the phase operator, and $\opn_i$ is the conjugate charge operator.
$\varphi_{\mathrm{ext},i}$ represents the external flux.
In this example, we will keep the fluxoniums at their sweetspots, or equivalently set $\varphi_{\mathrm{ext},i}=\pi$.
$H_{\mathrm{f},i}$ can be diagonalized
\begin{equation}
    H_{\mathrm{f},i} = \hbar \sum_l \omega_l \ket{l}\bra{l},
\end{equation}
and the transition frequencies are $\omega_{ab} = \omega_{b}-\omega_a$.
The coupling terms have the form
\begin{equation}
    H_{ij} = J_{\mathrm{C}} \opn_i \opn_j - J_{\mathrm{L}} \opv_i \opv_j,
\end{equation}
where $J_{\mathrm{C}}$ and $J_{\mathrm{L}}$ are the strength of capacitive and inductive coupling.
As a simplification, we choose to make the coupling strength constant and independent of the qubit pairs.
For certain reasonable parameters of a 2-fluxonium system, one can find $J_{\mathrm{C}}$ and $J_{\mathrm{L}}$ such that the two-body $ZZ$-crosstalk $c_{11} = 0$, where $c_{11}$ is defined in~\autoref{eq:zz_crosstalk_hamiltonian_general}.
We will apply time-dependent control through flux driving, which means that the operators $C_{i,k}$ in~\autoref{eq:single_qubit_control_hamiltonian} are $\opv_i$.

\begin{figure}
    \includegraphics[width=0.9\linewidth]{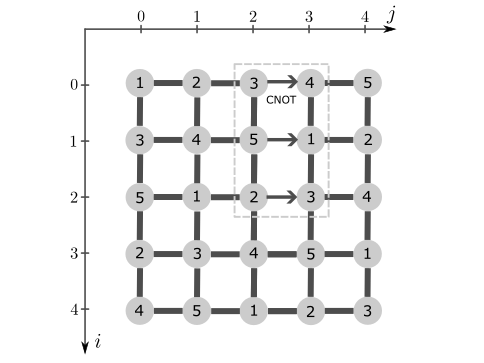}
    \caption{Frequency pattern and the simultaneous CNOT gates we simulated.
The frequency pattern is taken from~\cite{hertzbergLaserannealingJosephsonJunctions2021}.
We draw a $5\times 5$ region based on the frequency pattern.
For each row, the labels are repeating from 1 to 5 when going from left to right.
We also use the dashed box to indicate the 6 qubits which we perform simultaneous single and two-qubit gates on.
We label individual qubits by their coordinates $(i,j)$.
The 6-qubit region is then $\{Q_{ij}\}$ for $i\in \{0,1,2\}$ and $j\in \{2,3 \}$.
For the CNOT two-qubit gates, the arrows point from the control qubits to the target qubits.
We note that the label 3 appears twice in the 6-qubit region, which indicates that the parameters of the corresponding qubits are intended to be the same.
\label{fig:frequency_pattern_and_simul_cnot}}
\end{figure}

We consider the Hamiltonians for 6 fluxoniums arranged on a square lattice.
To ensure that the parameters of the fluxoniums are scalable for larger 2D processors, we arrange the parameters in a square lattice frequency pattern as described in~\cite{hertzbergLaserannealingJosephsonJunctions2021}.
This pattern is illustrated in~\autoref{fig:frequency_pattern_and_simul_cnot}, and the selected 6 qubits are enclosed by the dashed rectangle.
If the frequency pattern is exactly followed, then there should be five values ($i=1,\ldots ,5$) for each of the fluxonium parameters $E_{\mathrm{C},i}$, $E_{\mathrm{J},i}$, and $E_{\mathrm{L},i}$ in~\autoref{eq:hamiltonian_single_fluxonium}, where $i$ is the label of a site in~\autoref{fig:frequency_pattern_and_simul_cnot}.
The values of these parameters are approximately $E_{\mathrm{C},i} \approx 1$ GHz, $E_{\mathrm{J},i}\approx 4$ GHz, and $E_{\mathrm{L},i}\approx 1$ GHz, and the detailed values are provided in~\autoref{table:parameters_before_simple_optimization}.
However, as discussed in~\cite{hertzbergLaserannealingJosephsonJunctions2021}, it is not possible to fabricate superconducting qubits with the exact desired parameters.
The standard deviations can be a few percentages of the parameter values, which cannot be neglected.
Moreover, as discussed in~\cite{berkeTransmonPlatformQuantum2022}, a certain amount of disorder in the parameters can be helpful in making the computational qubits more localized and controllable.
Therefore, in our simulation, we will add randomness to the periodic layout of parameters.
More concretely, we set
\begin{equation}
    E_{\mathrm{C},i,xy} = E_{\mathrm{C},i} + \Delta E_{\mathrm{C},xy},
\end{equation}
where $E_{\mathrm{C},i,xy}$ is the $E_{\mathrm{C}}$ for the fluxonium with coordinate $(x,y)$ and site label $i$.
Each $\Delta E_{\mathrm{C},xy}$ is sampled from $0.01E_{\mathrm{C},i}\times\mathcal{N}(0,1)$, where $\mathcal{N}(0,1)$ is the normal distribution with mean 0 and variance 1.
In other words, we set the errors for fluxonium parameters to be around 1\%.
Similarly, we add random errors to $E_{\mathrm{J},i,xy}$ and $E_{\mathrm{L},i,xy}$.

For the simultaneous gate control schemes, we extend the single-qubit and two-qubit gate schemes in~\cite{nguyenBlueprintHighPerformanceFluxonium2022} to 6 fluxoniums by applying multiple control pulses simultaneously.
The control parameters are obtained through two optimizations.
First, we perform an optimization on an isolated 1-qubit or 2-qubit system, where we can choose initial values of control parameters using our understanding of the gate schemes.
Next, we use the results of the first optimization as initial values for the optimization of simultaneous gates on 6 fluxoniums.
It is likely that the initial values already lead to a reasonable performance of simultaneous gates, enabling smoother and faster local gradient optimization on the 6-fluxonium system.

The single-qubit gate scheme is the widely used Rabi $\pi$-rotation.
We apply a drive pulse on the control qubit $i$
\begin{equation}
    H_\epsilon(t) = \mathcal{E}(t) \cos(\omega_d t+\phi)  \opv_i, \label{eq:microwave_control_hamiltonian}
\end{equation}
where the pulse envelope $\mathcal{E}(t)$ is chosen to be
\begin{equation}
    \mathcal{E}(t) = \frac{\epsilon_d}{2}\left[ 1-\cos(2\pi t/\tau_{\mathrm{gate}}) \right].
    \label{eq:single_qubit_cosine_envelope}
\end{equation}
Due to the existence of higher energy levels and adjacent qubits, the optimal choice of $\omega_d$ is not $\omega_{01}$ of isolated fluxoniums.
We will also allow arbitrary $Z$-rotations before and after the gates since they can be essentially free~\cite{mckayEfficientGatesQuantum2017}.
The details of implementation are described in~\aref{sec:unitary_compensation_for_gates}.
The values of control parameters will be determined by the optimization procedure mentioned above.

The 2-qubit gate scheme we consider is the cross-resonance (CR) gate~\cite{rigettiFullyMicrowavetunableUniversal2010,degrootSelectiveDarkeningDegenerate2010}.
It is one of the 2-qubit gate schemes discussed in~\cite{nguyenBlueprintHighPerformanceFluxonium2022}.
To implement one CR gate, we need to apply microwave tone on one of the qubit at roughly the frequency $\omega_{10}$ of another qubit.
We will again consider drives of the form~\autoref{eq:microwave_control_hamiltonian} where the pulse envelope $\mathcal{E}(t)$ is chosen to be a flat-top pulse with cosine ramping
\begin{equation}
    \mathcal{E}(t) =
    \begin{cases}
        \frac{\epsilon_d}{2} (1-\cos \frac{\pi t}{\tau_{\mathrm{ramp}}}) ,& 0\leq t \leq \tau_{\mathrm{ramp}} \\
        \epsilon_d,     & \tau_{\mathrm{ramp}} \leq t \leq \tau_{\mathrm{gate}} -\tau_{\mathrm{ramp}} \\
        \frac{\epsilon_d}{2}(1-\cos \frac{\pi (\tau_{\mathrm{gate}}-t)}{\tau_{\mathrm{ramp}}}) ,& \tau_{\mathrm{gate}} -\tau_{\mathrm{ramp}} \leq t \leq \tau_{\mathrm{gate}}.
        \label{eq:cr_pulse_shape}
    \end{cases}
\end{equation}
By setting $\omega_d$ to a frequency close to the $0-1$ transition of the neighboring qubit, the resulted gate $U_\mathrm{CR}$ on the two qubits can be made to be closed to CNOT up to some single-qubit unitary operators.
Three CNOT gates can be performed simultaneously by applying drive pulses on 3 qubits with coordinates $(0,2), (1,2), (2,2)$ in~\autoref{fig:frequency_pattern_and_simul_cnot}.
For simplicity, we include errorless single-qubit gates before and after the simultaneous pulses to convert $U_\mathrm{CR}^{\otimes 3}$ into $\mathrm{CNOT}^{\otimes 3}$.
This assumption will lead to underestimates of the error rates for the two-qubit gates, as well as the logical error rates. If we wish to incorporate the errors associated with these compensations into the simulation, we must first determine how they are implemented in the syndrome measurement circuit shown in~\autoref{fig:syndrome_circuit}. For instance, it may be possible to combine a pair of compensations from consecutive two-qubit gates into a single compensation gate.

\subsection{Walsh-transform of idling Hamiltonian}
\label{sec:results_walsh_transform}

\begin{figure}
    \includegraphics[width=0.95\linewidth]{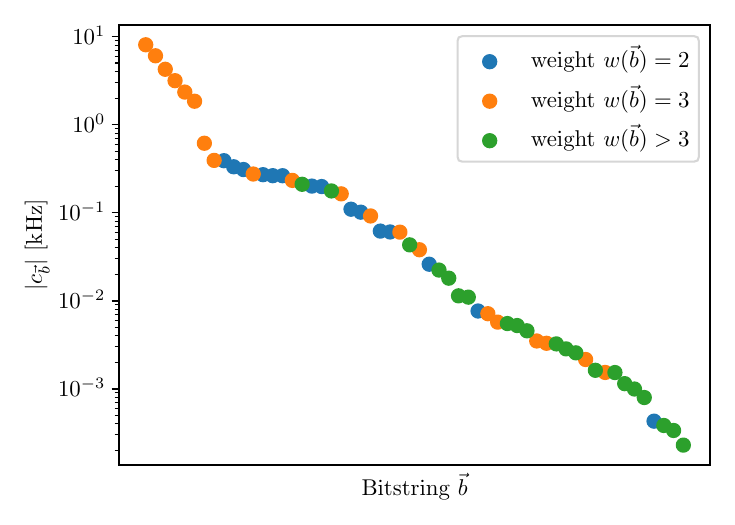}
    \caption{Walsh-transform analysis for the computational subspace of the Hamiltonian~\autoref{eq:two_local_fluxonium_Hamiltonian}.
Coefficients $c_{\vec{b}}$ are introduced in~\autoref{eq:zz_crosstalk_hamiltonian_general}.
The horizontal axis corresponds to different bitstrings $\vec{b}$.
We sort $|c_{\vec{b}}|$ by their magnitudes, and we can see that the largest  $|c_{\vec{b}}|$  correspond to $w(\vec{b})=3$.
We omit one point with $|c_{\vec{b}}|\approx 10^{-7}$ kHz and $w(\vec{b})>3$.
\label{fig:walsh_transform_6q}}
\end{figure}

First, we perform a Walsh-transform on the 6-fluxonium Hamiltonian $H(0)$ and plot the result in~\autoref{fig:walsh_transform_6q}.
To construct the Hamiltonian, we truncate each fluxonium qubit to the lowest 4 energy levels.
We observe that the largest coefficients of $c_{\vec{b}}$ in~\autoref{eq:zz_crosstalk_hamiltonian_general} correspond to $w(\vec{b})=3$.
This is consistent with the fact that the multi-path coupling scheme is capable of suppressing two-body $ZZ$ crosstalks~\cite{nguyenBlueprintHighPerformanceFluxonium2022}.
This also provides a concrete example for the discussion in~\autoref{sec:eigenvalue_crosstalk} and~\autoref{sec:perturbation_series}, illustrating that it is possible to reduce $c_{\vec{b}}$ with $w(\vec{b})=2$ without simultaneously reducing $c_{\vec{b}}$ with $w(\vec{b})=3$.
The values of $c_{\vec{b}}$ will not be used in calculating the logical error rate $\ler$, since there is no completely idling round and we do not assign correlated errors to idling and measurement operations.
However, this serves as a reminder that there are high-weight correlated errors even in the simplest operation.
We will see that high-weight correlated errors also dominate simultaneous gate operations in the absence of decoherence errors and with perfect control.

\subsection{Errors of single-qubit gates}
\label{sec:results_single_qubit_gate}

\begin{figure}
    \includegraphics[width=0.95\linewidth]{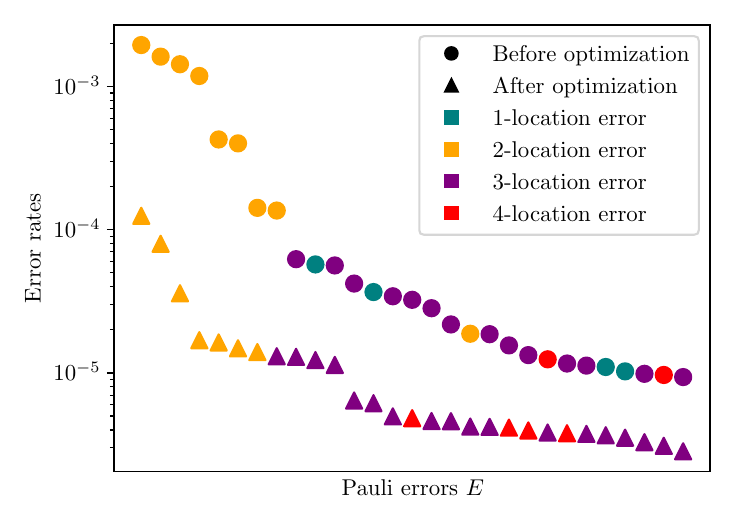}
    \caption{The largest Pauli error rates of simultaneous $X$ gates are shown both before and after optimization in~\autoref{sec:results_a_simple_gradient_optimization}.
Circle and triangle points represent the error rates before and after optimization, respectively, while colors indicate the number of locations affected by the errors.
The error rates are also listed in~\autoref{table:parameters_before_simple_optimization} and~\autoref{table:parameters_after_simple_optimization}.
We observe that errors with the highest probabilities are correlated errors affecting multiple qubits.
This is because decoherence errors have not been added and the control parameters have been optimized.
Further discussion on this can be found in~\autoref{sec:results_single_qubit_gate}.
\label{fig:simultaneous_1q_correlated_error}}
\end{figure}

To obtain the control parameters of simultaneous single-qubit $X$ gates, we follow the optimization procedure mentioned in~\autoref{sec:example_design_control_schemes}.
From now on, we truncate each fluxonium qubit to the lowest 3 energy levels.
The Pauli error rates converted from the unitary errors according to~\autoref{eq:error_prob_from_twirling} are illustrated in~\autoref{fig:simultaneous_1q_correlated_error} and listed in~\autoref{table:simultaneous_x_errors}, which we have not added decoherence errors yet.
We can see that the errors with largest probabilities are correlated errors on multiple qubits.
This is due to the fact that most single-qubit unitary errors can be corrected during the optimization of control parameters.
In greater detail, we know that on a single-qubit system, the rotation angles can be altered by tuning the amplitude and gate time in~\autoref{eq:single_qubit_cosine_envelope}.
The rotation axis can be adjusted by tuning the drive frequency $\omega_d$, phase $\phi$, and the $Z$-compensation.
As a result, if the decoherence error rates are small, the residual errors are predominantly multi-qubit.
This is different compared to the circuit-level depolarizing noise model.
We also see that most of the weight-2 correlated errors are on the pair $\mathrm{Q}_{02}$ and $\mathrm{Q}_{12}$.
The $0-1$ transition frequency difference between these two qubits is around 60Mhz (see~\autoref{fig:frequency_table} for the frequencies of qubits), which is the smallest among all neighboring pairs.
This might be the reason that this pair has the largest crosstalk error, although 60Mhz is still much larger compared to the 20Mhz frequency difference required by~\cite{nguyenBlueprintHighPerformanceFluxonium2022}.
In~\autoref{sec:results_a_simple_gradient_optimization}, we will see how gradient optimization can help us to reach a better frequency arrangement.

We can also compare the error rate distribution of simultaneous single-qubit gates shown in~\autoref{fig:simultaneous_1q_correlated_error} to the distribution of Walsh-transform coefficients depicted in~\autoref{fig:walsh_transform_6q}.
While both distributions share the feature that the dominant terms act on more qubits than some of the less dominant terms, the details differ significantly.
Specifically, the dominant terms of the Walsh-transform have support on three qubits, whereas the dominant terms of the simultaneous single-qubit gate have support on only two qubits.
So we can conclude that there is no direct correspondence between the two distributions.

\subsection{Errors of two-qubit gates}
\label{sec:results_two_qubit_gate}

For simultaneous CNOT gates, we can similarly compute and list the Pauli error rates in~\autoref{table:cnot_errors}.
We can see that the dominating errors are correlated errors acting on qubits from different gates.
The reason is similar to the case of single-qubit gate.
Here we allow arbitrary single-qubit compensation and optimize over the CR control parameters, which contains a similar number of parameters compared to the dimension of $\mathrm{SU}(4)$.

\subsection{One step of gradient optimization}
\label{sec:results_a_simple_gradient_optimization}

Since the simultaneous single-qubit gates are the dominant source of errors, we should prioritize addressing this issue.
Therefore, our first step is to perform a rough optimization aimed at mitigating this primary concern.
To assess the quality of the computed unitary $U_\mathrm{sim}$ of the simultaneous gate, a commonly used measure is the average fidelity (see~\autoref{eq:average_fidelity}).
Here we will compare the 6-qubit operators $U_\mathrm{sim,1q}$ and $U_\mathrm{sim,2q}$ to $X^{\otimes 6}$ and $\mathrm{CNOT}^{\otimes 3}$, respectively, in order to determine $F_{\mathrm{1q}}$ and $F_{\mathrm{2q}}$.
Given that the amount of leakage is negligible (see~\autoref{table:parameters_after_simple_optimization}), it is appropriate to use the average fidelity formula for unitary operators.
Let us consider the objective function:
\begin{equation}
O = \log_{10} (2 - F_{\mathrm{1q}} - F_{\mathrm{2q}}),
\label{eq:objective_fidelity}
\end{equation}
The gradient $\nabla O (\hamparam)$ is listed in~\autoref{table:parameters_before_simple_optimization}.
We will update the single fluxonium parameters $E_{\mathrm{C/J/L},i}$ based on this gradient.
However, we will not update the coupling strengths $J_\mathrm{C/L}$ at this stage, and we will optimize the control parameters separately after we make change to $E_{\mathrm{C/J/L},i}$.
This decision is motivated by the significant differences in suitable learning rates for each variable. We discuss the challenges associated with assigning learning rates to parameters with different units in~\aref{sec:optimization_discussion}.
Furthermore, as mentioned earlier, the high error rates observed in simultaneous single-qubit gates are likely attributed to the small frequency difference, which can be increased by modifying the fluxonium parameters alone. Consequently, we update $E_{\mathrm{C/J/L},i}$ according to the following equation:
\begin{equation}
    E_{\mathrm{C/J/L},i} \rightarrow E_{\mathrm{C/J/L},i} - 0.01 \mathrm{GHz^{2}}\times \frac{\partial O}{\partial E_{\mathrm{C/J/L},i}}.
\end{equation}
The units of $\sfrac{\partial O}{\partial E_{\mathrm{C/J/L},i}}$ are $\mathrm{GHz^{-1}}$ since we employ $\mathrm{GHz}$ as the unit for $E_{\mathrm{C/J/L},i}$ during the computation of $O$.
Following this, we conduct another optimization of the control parameters, resulting in the updated parameters listed in~\autoref{table:parameters_after_simple_optimization}.
Subsequently, we evaluate the errors in simultaneous single-qubit gates using the new parameters, as presented in~\autoref{table:simultaneous_x_errors_after_opt} and in~\autoref{fig:simultaneous_1q_correlated_error}.
Notably, the magnitudes of the correlated errors are considerably reduced compared to those in~\autoref{table:simultaneous_x_errors}.
This improvement can be directly attributed to the increased frequency difference between $Q_\mathrm{02}$ and $Q_\mathrm{12}$, as illustrated in~\autoref{fig:frequency_table}.
Consequently, we can deduce that the gradients $\sfrac{\partial O}{\partial E_{\mathrm{C/J/L},i}}$ encourage the parameters to exhibit a more favorable frequency pattern.

On the other hand, we have observed weight-4 correlated errors in~\autoref{fig:simultaneous_1q_correlated_error} and~\autoref{table:simultaneous_x_errors_after_opt} with a higher probability of $4.83\times 10^{-6}$ compared to several weight-3 errors. It is important to note that our LCPEM error model does not account for these weight-4 errors. Therefore, we should exercise caution when attempting to predict the behaviors of larger systems at low logical error rates.

\subsection{Logical error rates versus depolarizing rate}

\begin{figure}
    \includegraphics[width=\linewidth]{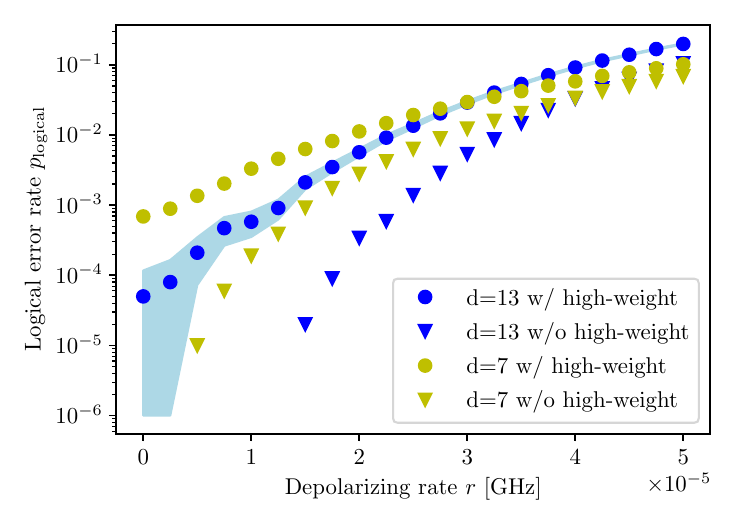}
    \caption{This figure depicts the relationship between logical-$X$ error rates $\ler$ and the decoherence rate after optimization done in~\autoref{sec:results_a_simple_gradient_optimization}.
    To provide a rough estimate, a value of $r=10^{-5}$ GHz corresponds to decoherence times $T_1$ and $T_2$ on the order of \SI{100}{\micro\second}.
The logical error rates $\ler$ are calculated for different depolarizing error rates $r$ and for two different surface code distances $d=7$ and $d=13$.
We also compute $\ler$ by setting the multi-gate correlated error rates $p_k^{(j)}=0$ for $k>1$.
Therefore, the triangle points can be viewed as the logical error rates if we completely ignore the high-weight errors in the simulation.
We can see that when the depolarizing rate $r$ gets smaller, the impact caused by high-weight errors becomes larger.
For each point, we perform the QEC simulation $10^5$ times to obtain the estimates of logical error rates.
The light blue colored region represents the values within three standard deviations of the mean corresponding to $d=13$ with multi-gate errors.
\label{fig:logical_error_vs_decoherence_after_opt}}
\end{figure}
The decoherence time and related errors are not determined from the Hamiltonian~\autoref{eq:2local_hamiltonian}.
There are also hopes that they can be continuously improved in the future because of better materials, fabrication processes and qubit designs, etc.
Therefore, we will treat the depolarizing rate $r$ in~\autoref{eq:add_decoherence_error_to_unitary} as a variable and plot the relation between the logical-$X$ error rates $\ler$ versus the strength of decoherence errors in~\autoref{fig:logical_error_vs_decoherence_after_opt}.
To provide a rough estimate, a value of $r=10^{-5}$ GHz corresponds to decoherence times $T_1$ and $T_2$ on the order of \SI{100}{\micro\second}.
This is consistent with the typical values observed in present-day superconducting qubits.
We can see that when $r$ gets smaller, the impact caused by high-weight errors becomes larger.
The breakeven values of $r$ for simulation with and without high-weight errors are roughly the same at $2\times 10^{-5}$.
This coincides with our argument that the high-weight errors will have less influence to the threshold.
However, $r$ is not the only noise model parameters, and other parameters such as the measurement error rate will have effects on the crossing points of the curves.

\subsection{Gradients of the logical error rates}

Let us now estimate the gradient $\nabla \ler (\hamparam)$ for parameters after optimization.
We set the depolarizing rate $r_{\mathrm{dep}} = 10^{-5} $ GHz and we consider the code distance $d=7$.
The values of parameters and corresponding elements in the gradient are listed in table~\autoref{table:parameters_after_simple_optimization}.
The estimated gradients $\sfrac{\Delta \ler}{\Delta p_k^{(j)}}$ are monotonically increasing with respect to the weight $k$.
This coincides with our intuition that high-weight correlated errors are more damaging for the logical error rates.
We have $\sfrac{\Delta \ler}{\Delta p_1^{(1)}} = 0$ because errors after Hadamard gates do not affect logical-$X$ error rates, as discussed in~\autoref{sec:qec_sim_workflow}.
After we have $\sfrac{\Delta \ler}{\Delta p_k^{(j)}}$, we can construct an intermediate objective function
\begin{equation}
    O(\hamparam) = \sum_{j,k} \frac{\Delta \ler}{\Delta p_k^{(j)}} p_k^{(j)}(\hamparam),
    \label{eq:objective_grad_logical_error_rates}
\end{equation}
and $\nabla O(\hamparam)\approx \nabla \ler (\hamparam)$.
We can then compute $\nabla O(\hamparam)$ with automatic differentiation library.

\section{Conclusion}

In this paper, we present a workflow for simulating logical error rates using the Hamiltonians of superconducting processors. Furthermore, the gradients of logical error rates can be efficiently computed through backpropagation with respect to the Hamiltonian parameters. This gradient information can then be utilized in optimization algorithms, such as gradient descent.

However, superconducting processors are highly intricate systems, and our approach does not completely address the challenge of accurately predicting the performance of quantum error correction. Consequently, despite the efficient gradient computation, the accuracy of the simulator remains a bottleneck. Among various issues, we specifically focus on spatially correlated coherent errors that arise from the always-on interactions in the Hamiltonians. These errors can significantly impact the logical error rates and even pose a threat to the threshold theorem. We provide arguments and an example to illustrate why even well-designed coupling scheme, such as tunable couplers, fail to eliminate all correlated errors. Therefore, it is crucial to consider such errors when evaluating the sub-threshold performance of a fault-tolerant scheme for superconducting processors. In this study, we compute these errors by performing exact time-evolution on a small number of qubits. Although this is insufficient for proving the threshold theorem, it adequately demonstrates the influence of spatially correlated errors.

There are several important details that we have not included in our simulation workflow. To mention a few, we have not simulated the readout resonators and possible correlated measurement errors. Additionally, we have not considered the decoherence mechanisms that may depend on the parameters of the qubits and resonators. However, it appears feasible to incorporate these factors incrementally into the workflow.
There is another factor that we have omitted, which could pose a significant challenge for simulation.
In superconducting processor experiments, a certain degree of calibration or black-box optimization is typically performed.
Calibration is necessary due to the inherent difficulty in achieving a perfect description of the quantum processor before conducting experiments.
Unpredictable factors, such as two-level system defects, can fluctuate randomly in frequency and time, thereby impacting the coherence of qubits~\cite{klimovFluctuationsEnergyRelaxationTimes2018}.
Our workflow does not account for the uncertainties inherent in the processors and the calibration process, which can undeniably increase the complexity of accurate simulation.
One potential solution to address these issues involves simulating average error rates resulting from calibration by repeatedly sampling the device with its variations.
An illustrative example of computing such average error rates can be found in~\cite{niIntegratingQuantumProcessor2022}.

\section{Acknowledgement}

The authors would like to thank Barbara Terhal, Fang Zhang, Hui-Hai Zhao and Feng Wu for constructive discussions. This work was supported by Alibaba Group through Alibaba Research Intern Program, and conducted when Z.~W. was a research intern at Alibaba Group.
\bibliographystyle{apsrev4-1}
\bibliography{bib}

\appendix

\section{Syndrome extraction circuit}

\begin{figure*}
    \includegraphics[width=0.9\linewidth]{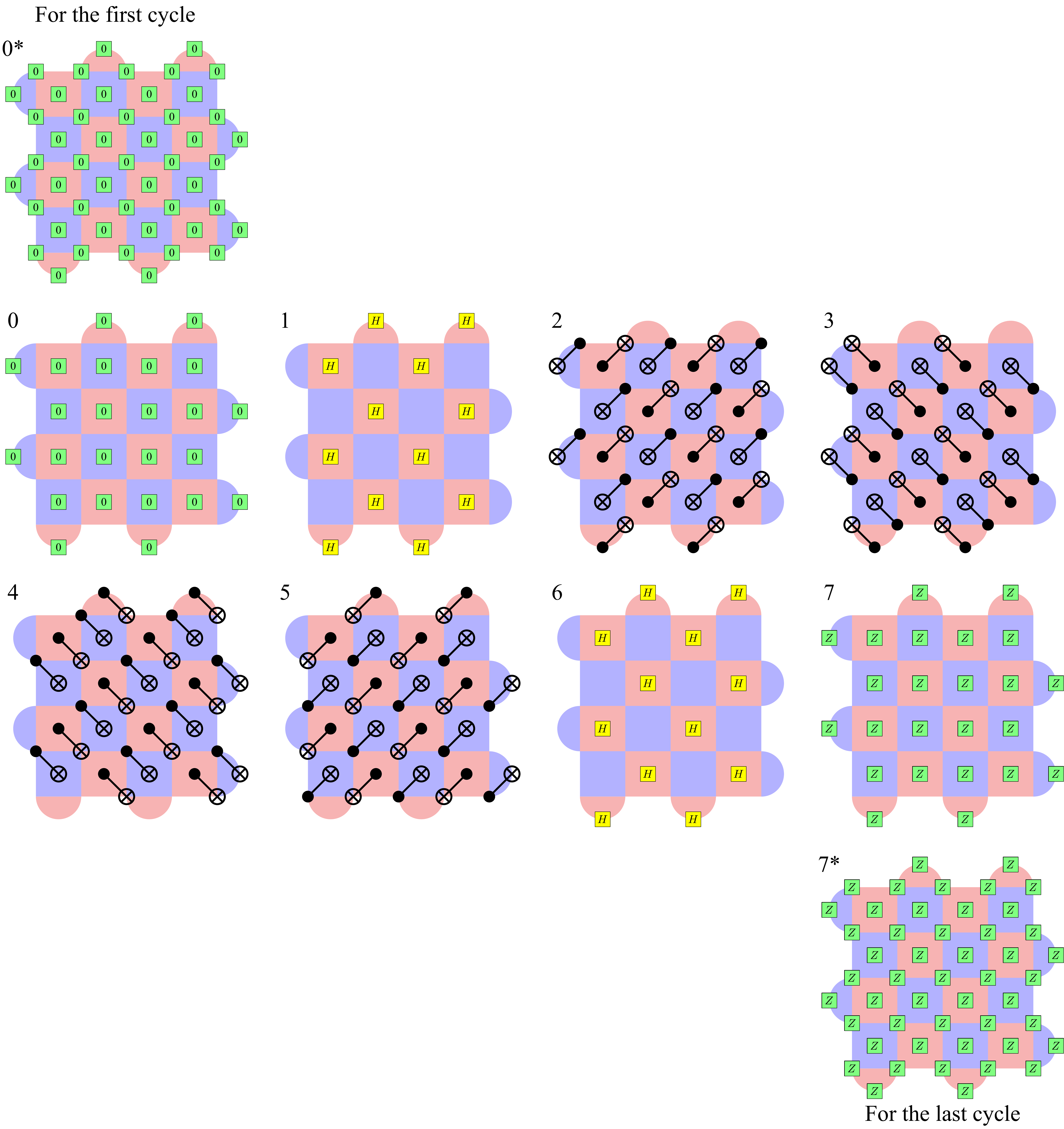}
    \caption{Syndrome measurement circuit for surface code with CNOT gates being the 2-qubit gates.
    There is a data qubit on each vertex and an ancilla qubit on each plaquette.
    Operation labeled by 0 is the state preparation in $\ket{0}$ state.
$H$ and CNOT are the corresponding single and two-qubit gates.
$Z$ is the measurement in the computational basis.
Qubits without operations are idling during that time step.
The time duration of the idling depends on the duration of other operations in the same time step.
The first and last QEC cycles have some different steps, which is marked with asterisks.
\label{fig:syndrome_circuit}}
\end{figure*}

In~\autoref{fig:syndrome_circuit}, we show the syndrome extraction circuit used in our simulation.
There are qubits on the center and vertices of plaquettes.
Operation labeled by 0 is the state preparation in $\ket{0}$ state.
H and CNOT are the corresponding single and two-qubit gates.
$Z$ is the measurement in the computational basis.
The first and last QEC cycles have some different steps.

\section{Challenges in Optimization}
\label{sec:optimization_discussion}

Most of the optimization performed in this study focuses on control parameters, resulting in solutions with low error rates, as demonstrated by~\autoref{fig:simultaneous_1q_correlated_error} and the tables in~\aref{sec:data_details}.
In~\autoref{sec:results_a_simple_gradient_optimization}, we execute a single step of gradient optimization for qubit parameters, manually selecting an appropriate step size.
However, we refrain from conducting extensive optimization involving both device and control parameters due to the need for further improvements to enhance optimization efficiency and stability.

Firstly, there is freedom in choosing the units of the parameters for which the gradients are computed, and this can affect the choice of learning rates and the direction of gradients.
For instance, the update rule for gradient descent is expressed by the equation:
\begin{equation}
a_{n+1} = a_{n} - r \frac{\partial O}{\partial a}.
\end{equation}
Given that the objective function $O$ is typically dimensionless, we can infer that the step size $r$ possesses units.
If we change the unit of $a$, the value of $r$ need to be changed accordingly to maintain the same update rule.
This also implies that assigning the same learning rate to all parameters in the optimization is not ideal.
Additionally, consider a simple objective function
\begin{equation}
    f(a_{\mathrm{Hz}},b_{\mathrm{s}}) = a_{\mathrm{Hz}} \ \mathrm{Hz} \times b_{\mathrm{s}} \ \mathrm{s}.
\end{equation}
If we change the units to GHz and ns, then the same function would still be
\begin{equation}
    g(a_{\mathrm{GHz}},b_{\mathrm{ns}}) = a_{\mathrm{GHz}} \ \mathrm{GHz} \times b_{\mathrm{ns}} \ \mathrm{ns}.
\end{equation}
However, for the derivatives we have
\begin{equation}
    \frac{\partial g}{\partial a_{\mathrm{GHz}}} = b_{\mathrm{ns}} = 10^9 \times \frac{\partial f}{\partial a_{\mathrm{Hz}}},
\end{equation}
\begin{equation}
    \frac{\partial g}{\partial b_{\mathrm{ns}}}=a_{\mathrm{GHz}} = 10^{-9} \times \frac{\partial f}{\partial b_{\mathrm{s}}}.
\end{equation}
On the other hand, $a_{\mathrm{GHz}} = 10^{-9} a_{\mathrm{Hz}} $, $b_{\mathrm{ns}} = 10^9 b_{\mathrm{s}}$.
So the direction of the gradient changes drastically after we change the units.
To mitigate these challenges, we should evaluate second-order derivatives and use them to set the step sizes periodically during the optimization process.

As usual, the optimization of device and control parameters will generally have numerous local minima.
This is already the case even if we only consider the optimization of control parameters.
One cause of the multiple local minima is that unitary rotations are periodic with respect to the rotation angle.
For example, if we want to perform an $X$ gate, the rotation angles can be $\pi + 2n\pi$.
These angles are distinct, and the control parameters to realize these rotation angles will form local minima.
Similarly, we have $\mathrm{CNOT}^{2n+1}=\mathrm{CNOT}$ and $\mathrm{CZ}^{2n+1} = \mathrm{CZ}$.
In general, we expect that $n=0$ will correspond to operations with the lowest error rates.
However, this is not necessarily true.
For example, an $X$ gate with a large $n$ might have a lower dephasing error or $ZZ$-crosstalk error due to having a larger drive pulse amplitude~\cite{guoDephasingInsensitiveQuantumInformation2018}.
Additionally, there are local minima that have other origins.
For instance, we can see in figures 4 and 7 of~\cite{nguyenBlueprintHighPerformanceFluxonium2022} that the optimal fidelities can oscillate when we continuously vary certain other control or device parameters.
Moreover, qubit frequency arrangement on a processor also likely involves optimization of discrete parameters such as the order of frequencies.
The above discussion implies that we cannot expect to find the global minimum with a single run of gradient optimization.
On the other hand, we can just start with multiple initial parameters and the gradient optimization can help us find different local minima faster.

\section{Fidelity and high weight errors}
\label{sec:fidelity_high_weight_errors}
It is known that the average fidelities of gate operations are not enough to predict their performance in the syndrome extraction circuits.
In particular, leakage and coherent errors can be much more damaging in the worst-case scenario compared to stochastic Pauli errors.
But the average fidelity does not assign correct weights to these error components.
In fact, it is suggested in~\cite{iyerSmallQuantumComputer2018} that there is no way to find a few critical parameters to predict the impact of a noisy operation on QEC.

Considering that spatially correlated errors are a significant focal point in the main text, we aim to specifically investigate how the widely utilized metric, average fidelity, responds to these types of errors.
For that, let us consider a simple stochastic Pauli error channel on $m$ qubits:
\begin{equation}
    \mathcal{N}(\ket{\psi})=
\begin{cases}
    Z^{\otimes n}\ket{\psi} ,& \text{with probability} \  p\\
    \ket{\psi},              & \text{otherwise,}
\end{cases}
\end{equation}
where $n\leq m$ is the weight of the error $Z^{\otimes n}$.
There are two very different ways to compute the fidelity metric of such noise channels.
Assuming that the noise channel occurs after a single qubit gate on qubit $0$ and that the system is initially in a product state, we can first trace out the other ``environment'' qubits and obtain a noise channel solely on qubit $0$.
The resulted single-qubit channel is simply
\begin{equation}
    \mathcal{N}'(\ket{\psi})=
\begin{cases}
    Z_0\ket{\psi} ,& \text{with probability} \  p\\
    \ket{\psi},              & \text{otherwise.}
\end{cases}
\end{equation}
Therefore, it is clear that we have already lost track of the weight $n$ of the Pauli error $Z^{\otimes n}$, and no fidelity metric of $\mathcal{N}'$ can depend on $n$.
The other way to compute the fidelity metric is directly comparing $\mathcal{N}$ with identity channel $\mathcal{I}$.
We will use the formula in~\cite{nielsenSimpleFormulaAverage2002}
\begin{equation}
    \bar{F} (\mathcal{E},I) = \frac{\sum_j \tr (U_j^\dagger \mathcal{E}(U_j)) + d^2}{d^2(d+1)}, \label{eq:average_fidelity}
\end{equation}
where $d$ is the dimension of the system and the unitary operators $U_j$ forms an orthonormal operator basis, i.e.
$\tr(U_j^\dagger U_k) =\delta_{jk} d$.
Here we can choose the orthonormal basis to be the $m$-qubit Pauli group $\{P_j\}$, where $P_j$ is tensor product of single-qubit Pauli matrices.
To show that $\bar{F} (\mathcal{N},I)$ does not depends on $n$, we only need to show $\sum_j \tr (P_j^\dagger \mathcal{N}(P_j))$ does not.
We notice
\begin{equation}
    P_j^\dagger \mathcal{N}(P_j)=
\begin{cases}
    I ,& \text{if} \   P_j \text{ commutes with} \ Z^{\otimes n}\\
    (1-2p) I            & \text{otherwise.}
\end{cases}
\end{equation}
We know that for any $P_j \neq I$, half of the $m$-qubit Pauli group will commute with $P_j$.
In particular, there are always same amount of $P_j$ commute and anti-commute with $Z^{\otimes n}$ in the summation $\sum_j \tr (P_j^\dagger \mathcal{N}(P_j))$.
Therefore, $\bar{F} (\mathcal{N},I)$ does not depends on $n$.

\section{Details of time-evolution simulation}
\label{sec:detail_time-evolution simulation}

In this work, we only compute the unitary time-evolution according to Schrödinger's equation, which is enough to demonstrate the existence of correlated errors.
However, in some cases, it is crucial to simulate the evolution of the density matrix using the master equation.
For instance, if a perfect 2-qubit gate $U_2=U(T,0)$ exists, but the partial time evolution $U(t,0)$ for $t<T$ acts non-trivially on $k>2$ qubits, a decoherence error occurring at time $t$ will generally spread to a weight-$k'$ error at time $T$, where $k' \geq k$.
Similarly, if $U(t,0)$ maps computational states to non-computational states, a non-leakage error such as dephasing at time $t$ can be transformed into a leakage error at time $T$.
These transformations of errors can drastically change their impact on logical error rates, and thus, we cannot simply add the decoherence errors to unitary time-evolutions in the above cases.

The number of qubits included in the numerical simulation can be chosen based on the computing resource constraint.
Obviously, if time permits, we want to simulate as many qubits as possible while maintaining a reasonable time-evolution simulation accuracy.
In this work, we simulate $N=6$ superconducting qubits while keeping 3 to 4 levels for each qubit.
We aim to perform one simulation run in a short time because we need to perform optimizations that require many repetitions of the simulation.

\subsection{Step by step computation}

The detailed step-by-step computation can be summarized as follows:

\begin{enumerate}

    \item For the idling Hamiltonian $H^{(1)}_{i} (0)$ of each qubit, calculate its eigenvalues and eigenbasis (see~\cite{groszkowskiScqubitsPythonPackage2021} for examples).
In the following steps, we will write operators $H^{(1)}_{i} (0)$, $\hat n_i$ and $ \hat \phi_i$ in the energy eigenbasis of the corresponding qubits.
Note that we are still in a product basis.
Let us call this specific product basis $B_\mathrm{bare}$.

    \item Construct the Hamiltonian $H(0)$ from~\autoref{eq:2local_hamiltonian} in the product basis $B_\mathrm{bare}$.

    \item Find the computational states $\{\ket{\psi_{\vec{b}}}\}$ according to the procedure in~\autoref{sec:convert_unitary_error}.
The states are written in the product basis $B_\mathrm{bare}$.

    \item We want to compute the time evolution for initial states in $\{\ket{\psi_{\vec{b}}}\}$.
We discretize the time evolution with step size $\delta t$.
For each time step, the new state is calculated by $|\psi(t+\delta t)\rangle = U(t+\delta t, t)|\psi(t)\rangle$, where $U(t+\delta t, t) = e^{-iH(t)\delta t}$ is the propagator operator.

    \item Apply the Suzuki-Trotter decomposition (STD) to the composited system Hamiltonian $e^{-iH(t)\delta t}$, and we could get a set of local Hamiltonian.
In the first-order STD, we have
    \begin{equation}
        e^{-iH(t)\delta t} \approx \prod_i e^{-iH^{(1)}_{i}(t)\delta t} \prod_{i,j} e^{-iH^{(2)}_{ij}\delta t}
    \end{equation}
    The local hamiltonians on the right hand side act on one or two qubits, and therefore we can easily perform matrix exponentiations and multiply them with $|\psi(t)\rangle$.
    The time step sizes used are $\delta t = 0.02$ ns for simultaneous single-qubit gates and $\delta t = 0.065$ ns for simultaneous two-qubit gates. It is worth noting that these step sizes may not be sufficiently small for the computation to converge. Nevertheless, the results obtained are qualitatively similar when compared to those obtained using significantly smaller step sizes.
    \item For initial state $\ket{\psi_{\vec{b}}}$, let us call the final state  $\ket{\varphi_{\vec{b}}}_{\mathrm{final}}$.
    We can expand the final state in the incomplete basis $\{\ket{\psi_{\vec{b}}}\}$
    \begin{equation}
        \ket{\varphi_{\vec{b}_i}}_{\mathrm{final}} = \sum_{\vec{b}_j} c_{ij} \ket{\psi_{\vec{b}_j}} + l_i \ket{\psi_{\perp}}.
        \label{eq:time_evo_result_and_leakage}
    \end{equation}
    Coefficients $c_{ij}$ form a square matrix $C$.
If the leakages $l_i$ are all small, then $C$ is approximately a unitary matrix.
This is the case for our simultaneous gates in~\autoref{eq:time_evo_result_and_leakage}, and for the following steps we will assume $C = U_{\mathrm{sim}}$ is unitary.

    \item Apply unitary compensation to $U_{\mathrm{sim}}$ by the procedure described in~\autoref{sec:unitary_compensation_for_gates}.

    \item Let $U_\mathrm{error} = U_{\mathrm{sim}}^{\dagger} U_{\mathrm{target}}$.
    We can expand $U_\mathrm{error}$ in the Pauli basis $U_\mathrm{error} = \sum_{\vec{j}} a(\vec{j})P_{\vec{j}}$ according to~\autoref{sec:convert_unitary_error}.

    \item We extract LCPEM parameters from $a(\vec{j})^2$.
For the simultaneous single-qubit gates, we can compute
    \begin{equation}
        p^{(i)}_k = \frac{1}{n^{(i)}_k}\sum_{\vec{j}\in W^{(i)}_k} a(\vec{j})^2  ,
        \label{eq:extract_1q_gate_p1k}
    \end{equation}
    where $W^{(i)}_k$ is the set of $\vec{j}$ with nonzero elements on  $k$  neighboring gates and $n^{(i)}_k$ is the number of sets of $k$ neighboring gate locations in the simulated region.
For the 6-qubit square lattice in~\autoref{fig:local_correlated_noise_model}, we have $n^{(1)}_2 =7 $ and $n^{(1)}_3 = 10$, $n^{(2)}_2 =2 $ and $n^{(2)}_3 = 1$.
We add decoherence noise to $p^{(i)}_1$ according to~\autoref{eq:add_decoherence_error_to_unitary}.

    \item We perform standard Clifford circuit simulation with above noise model and standard decoding with union-find decoder~\cite{huangFaulttolerantWeightedUnionfind2020}.
This process is repeated many times to estimate the logical error rates $\ler$.
\end{enumerate}

\section{Unitary compensation for single and two-qubit gates}
\label{sec:unitary_compensation_for_gates}

As we mentioned in~\autoref{sec:example_design_control_schemes}, we allow compensation to the simultaneous single and two-qubit gates.
For single-qubit gates, we apply $Z$-rotations $Z(\vec{\theta}_{\mathrm{bef}})$ and $Z(\vec{\theta}_{\mathrm{aft}})$ before and after the gates, where
\begin{equation}
    Z(\vec{\theta}) = \bigotimes_i Z(\theta_i).
\end{equation}
In the numerical example considered in the work, we have $i\in [1,6]$.
For two-qubit gates, we apply single-qubit unitaries $U_{i, \mathrm{bef}}$ and $U_{i, \mathrm{aft}}$ before and after the gates on qubit $i$.
Each unitary operation is parameterized 3 rotation angles.

We obtain these compensation operations through optimization.
In this work, they are optimized together with the Hamiltonian parameters $\hamparam$.

\section{Miscellaneous Data} 
\label{sec:data_details}

\renewcommand{\arraystretch}{1.5}%
\begin{table*}
    \small
    \centering
    \resizebox{0.55\textwidth}{!}{%
    \begin{tabular}{| l | l | r || l | l | r |}%
        \hline%
        \multicolumn{3}{|c||}{Device parameters}&\multicolumn{3}{c|}{Control parameters}\\%
        \hline%
        Parameter&Value (GHz)&Gradient&Parameter&Value&Gradient\\%
        \hline%
        $E_{C, 3}$&1.000e+00&{-}1.116e+00&$\epsilon_{{d}}$($\mathrm{Q}_{02}$)&1.184e{-}02 GHz&3.907e{-}01 \\%
        $E_{J, 3}$&4.000e+00&3.573e{-}01&$\tau_{\text{gate}}$($\mathrm{Q}_{02}$)&4.006e+01 ns&{-}1.257e{-}01 \\%
        $E_{L, 3}$&1.000e+00&{-}9.811e{-}01&$\omega_{{d}}/2 \pi$($\mathrm{Q}_{02}$)&5.708e{-}01 GHz&{-}4.038e{-}02 \\%
        $E_{C, 4}$&1.000e+00&4.461e{-}01&$\epsilon_{{d}}$($\mathrm{Q}_{03}$)&1.085e{-}02 GHz&9.047e{-}02 \\%
        $E_{J, 4}$&4.000e+00&{-}1.332e{-}01&$\tau_{\text{gate}}$($\mathrm{Q}_{03}$)&4.004e+01 ns&7.753e{-}04 \\%
        $E_{L, 4}$&8.000e{-}01&4.058e{-}01&$\omega_{{d}}/2 \pi$($\mathrm{Q}_{03}$)&4.155e{-}01 GHz&{-}1.235e{-}01 \\%
        $E_{C, 5}$&1.000e+00&9.533e{-}01&$\epsilon_{{d}}$($\mathrm{Q}_{12}$)&1.141e{-}02 GHz&{-}1.364e{-}01 \\%
        $E_{J, 5}$&4.000e+00&{-}2.952e{-}01&$\tau_{\text{gate}}$($\mathrm{Q}_{12}$)&4.005e+01 ns&{-}2.826e{-}02 \\%
        $E_{L, 5}$&9.000e{-}01&8.537e{-}01&$\omega_{{d}}/2 \pi$($\mathrm{Q}_{12}$)&5.048e{-}01 GHz&3.319e{-}02 \\%
        $E_{C, 1}$&1.000e+00&{-}5.636e{-}02&$\epsilon_{{d}}$($\mathrm{Q}_{13}$)&1.227e{-}02 GHz&{-}3.663e{-}01 \\%
        $E_{J, 1}$&4.000e+00&1.697e{-}02&$\tau_{\text{gate}}$($\mathrm{Q}_{13}$)&4.004e+01 ns&{-}5.045e{-}03 \\%
        $E_{L, 1}$&1.100e+00&{-}4.611e{-}02&$\omega_{{d}}/2 \pi$($\mathrm{Q}_{13}$)&6.678e{-}01 GHz&{-}2.309e{-}02 \\%
        $E_{C, 2}$&1.000e+00&4.883e{-}03&$\epsilon_{{d}}$($\mathrm{Q}_{22}$)&1.274e{-}02 GHz&7.498e{-}02 \\%
        $E_{J, 2}$&4.000e+00&{-}3.898e{-}03&$\tau_{\text{gate}}$($\mathrm{Q}_{22}$)&4.004e+01 ns&8.422e{-}04 \\%
        $E_{L, 2}$&1.200e+00&8.106e{-}03&$\omega_{{d}}/2 \pi$($\mathrm{Q}_{22}$)&7.963e{-}01 GHz&{-}2.661e{-}02 \\%
        $J_{C}$&1.150e{-}02&{-}3.680e{-}02&$\epsilon_{{d}}$($\mathrm{Q}_{23}$)&1.167e{-}02 GHz&{-}3.260e{-}01 \\%
        $J_{L}$&{-}2.000e{-}03&{-}9.274e+00&$\tau_{\text{gate}}$($\mathrm{Q}_{23}$)&4.004e+01 ns&{-}3.975e{-}03 \\%
        &&&$\omega_{{d}}/2 \pi$($\mathrm{Q}_{23}$)&5.591e{-}01 GHz&{-}1.657e{-}02 \\%
        \hline
        &&&$\epsilon_{{d}}$($\mathrm{Q}_{02}$)&3.082e{-}02 GHz&1.747e{-}02 \\%
        &&&$\tau_{\text{ramp}}$($\mathrm{Q}_{02}$)&3.020e+01 ns&{-}1.019e{-}02 \\%
        &&&$\tau_{\text{plateau}}$($\mathrm{Q}_{02}$)&7.000e+01 ns&{-}5.265e{-}03 \\%
        &&&$\omega_{{d}}/2 \pi$($\mathrm{Q}_{02}$)&4.191e{-}01 GHz&{-}5.185e{-}01 \\%
        &&&$\epsilon_{{d}}$($\mathrm{Q}_{12}$)&2.882e{-}02 GHz&2.379e{-}02 \\%
        &&&$\tau_{\text{ramp}}$($\mathrm{Q}_{12}$)&2.998e+01 ns&1.615e{-}04 \\%
        &&&$\tau_{\text{plateau}}$($\mathrm{Q}_{12}$)&7.000e+01 ns&2.176e{-}04 \\%
        &&&$\omega_{{d}}/2 \pi$($\mathrm{Q}_{12}$)&6.655e{-}01 GHz&3.412e{-}02 \\%
        &&&$\epsilon_{{d}}$($\mathrm{Q}_{22}$)&3.390e{-}02 GHz&{-}4.356e{-}02 \\%
        &&&$\tau_{\text{ramp}}$($\mathrm{Q}_{22}$)&3.007e+01 ns&{-}5.547e{-}04 \\%
        &&&$\tau_{\text{plateau}}$($\mathrm{Q}_{22}$)&6.994e+01 ns&{-}4.577e{-}04 \\%
        &&&$\omega_{{d}}/2 \pi$($\mathrm{Q}_{22}$)&5.592e{-}01 GHz&5.530e{-}02 \\%

        \hline%
        \end{tabular}}
    \caption{\label{table:parameters_before_simple_optimization} The table contains device and control parameters  before the optimization conducted in~\autoref{sec:results_a_simple_gradient_optimization}.
The gradient listed is $\nabla O (\hamparam)$, where $O$ is the objective function containing both single and two-qubit gates' fidelities defined in~\autoref{eq:objective_fidelity}.
The units of gradients are the reciprocal of the units of their corresponding values.}
\end{table*}

\normalsize%
\renewcommand{\arraystretch}{1.5}%
\begin{table*}
    \centering
    \resizebox{0.5\textwidth}{!}{%
    \begin{tabular}{| l | l | r || l | l | r |}%
        \hline%
        \multicolumn{3}{|c||}{Device parameters}&\multicolumn{3}{c|}{Control parameters}\\%
        \hline%
        Parameter&Value (GHz)&Gradient&Parameter&Value&Gradient\\%
        \hline%
        $E_{C, 3}$&1.009e+00&{-}3.214e{-}03&$\epsilon_{{d}}$($\mathrm{Q}_{02}$)&1.189e{-}02 GHz&1.529e{-}03 \\%
        $E_{J, 3}$&3.987e+00&1.094e{-}03&$t_{\text{single}}$($\mathrm{Q}_{02}$)&3.982e+01 ns&7.183e{-}07 \\%
        $E_{L, 3}$&1.007e+00&{-}3.040e{-}03&$\omega_{{d}}/2 \pi$($\mathrm{Q}_{02}$)&5.901e{-}01 GHz&{-}8.580e{-}05 \\%
        $E_{C, 4}$&9.912e{-}01&8.710e{-}04&$\epsilon_{{d}}$($\mathrm{Q}_{03}$)&1.089e{-}02 GHz&1.247e{-}03 \\%
        $E_{J, 4}$&3.984e+00&{-}2.464e{-}04&$t_{\text{single}}$($\mathrm{Q}_{03}$)&3.981e+01 ns&2.294e{-}06 \\%
        $E_{L, 4}$&7.925e{-}01&7.332e{-}04&$\omega_{{d}}/2 \pi$($\mathrm{Q}_{03}$)&4.089e{-}01 GHz&1.264e{-}04 \\%
        $E_{C, 5}$&9.895e{-}01&{-}1.424e{-}03&$\epsilon_{{d}}$($\mathrm{Q}_{12}$)&1.137e{-}02 GHz&1.966e{-}03 \\%
        $E_{J, 5}$&3.999e+00&4.775e{-}04&$t_{\text{single}}$($\mathrm{Q}_{12}$)&3.985e+01 ns&1.116e{-}06 \\%
        $E_{L, 5}$&8.906e{-}01&{-}1.432e{-}03&$\omega_{{d}}/2 \pi$($\mathrm{Q}_{12}$)&4.895e{-}01 GHz&6.952e{-}05 \\%
        $E_{C, 1}$&9.932e{-}01&1.340e{-}04&$\epsilon_{{d}}$($\mathrm{Q}_{13}$)&1.236e{-}02 GHz&2.102e{-}03 \\%
        $E_{J, 1}$&3.970e+00&{-}2.963e{-}05&$t_{\text{single}}$($\mathrm{Q}_{13}$)&3.985e+01 ns&5.131e{-}06 \\%
        $E_{L, 1}$&1.092e+00&7.108e{-}05&$\omega_{{d}}/2 \pi$($\mathrm{Q}_{13}$)&6.689e{-}01 GHz&{-}3.793e{-}04 \\%
        $E_{C, 2}$&1.014e+00&{-}3.001e{-}03&$\epsilon_{{d}}$($\mathrm{Q}_{22}$)&1.282e{-}02 GHz&6.141e{-}05 \\%
        $E_{J, 2}$&4.058e+00&1.003e{-}03&$t_{\text{single}}$($\mathrm{Q}_{22}$)&3.980e+01 ns&4.030e{-}08 \\%
        $E_{L, 2}$&1.217e+00&{-}2.576e{-}03&$\omega_{{d}}/2 \pi$($\mathrm{Q}_{22}$)&7.962e{-}01 GHz&1.744e{-}07 \\%
        $J_{C}$&1.150e{-}02&{-}1.335e{-}03&$\epsilon_{{d}}$($\mathrm{Q}_{23}$)&1.181e{-}02 GHz&1.286e{-}03 \\%
        $J_{L}$&{-}2.000e{-}03&{-}3.099e{-}01&$t_{\text{single}}$($\mathrm{Q}_{23}$)&3.985e+01 ns&4.132e{-}06 \\%
        &&&$\omega_{{d}}/2 \pi$($\mathrm{Q}_{23}$)&5.787e{-}01 GHz&2.280e{-}04 \\%

        \hline%
        &&&$\epsilon_{{d}}$($\mathrm{Q}_{02}$)&2.885e{-}02 GHz&{-}9.583e{-}04 \\%
        &&&$t_{\text{ramp}}$($\mathrm{Q}_{02}$)&2.993e+01 ns&{-}2.868e{-}05 \\%
        &&&$t_{\text{plateau}}$($\mathrm{Q}_{02}$)&6.994e+01 ns&{-}3.455e{-}05 \\%
        &&&$\omega_{{d}}/2 \pi$($\mathrm{Q}_{02}$)&4.121e{-}01 GHz&{-}4.976e{-}03 \\%
        &&&$\epsilon_{{d}}$($\mathrm{Q}_{12}$)&2.844e{-}02 GHz&1.208e{-}03 \\%
        &&&$t_{\text{ramp}}$($\mathrm{Q}_{12}$)&2.995e+01 ns&{-}3.291e{-}05 \\%
        &&&$t_{\text{plateau}}$($\mathrm{Q}_{12}$)&6.995e+01 ns&{-}8.484e{-}06 \\%
        &&&$\omega_{{d}}/2 \pi$($\mathrm{Q}_{12}$)&6.684e{-}01 GHz&2.617e{-}03 \\%
        &&&$\epsilon_{{d}}$($\mathrm{Q}_{22}$)&3.720e{-}02 GHz&6.181e{-}04 \\%
        &&&$t_{\text{ramp}}$($\mathrm{Q}_{22}$)&2.999e+01 ns&{-}3.514e{-}04 \\%
        &&&$t_{\text{plateau}}$($\mathrm{Q}_{22}$)&6.996e+01 ns&{-}1.452e{-}04 \\%
        &&&$\omega_{{d}}/2 \pi$($\mathrm{Q}_{22}$)&5.809e{-}01 GHz&5.391e{-}03 \\%

        \hline%
        \multicolumn{3}{|c||}{Single-qubit gate error}&\multicolumn{3}{c|}{Two-qubit gate error}\\%
        \hline%
        Type&Error probability&Gradient&Type&Error probability&Gradient\\%
        \hline%
        $p^{(1)}_1$ &8.419e-08&0.000e0&$p^{(2)}_1$ &7.390e-06&4.220e0 \\%
        $p^{(1)}_2$ &4.466e-5&3.359e0&$p^{(2)}_2$ &2.501e-4&9.755e0 \\%
        $p^{(1)}_3$ &1.706e-5&2.403e1&$p^{(2)}_3$ &1.088e-4&1.801e1 \\%
        $p_{\mathrm{leak}}$ & 2.747e-6 & & $p_{\mathrm{leak}}$ & 3.171e-5 &  \\%
        \hline%
        \multicolumn{3}{|c||}{Spam error}&\multicolumn{3}{c|}{Depolarizing rate}\\%
        \hline
        $p_\mathrm{reset}$ & 5.000e-3 & 7.460e-1& $r_\mathrm{dep}$ & 1.000e-5 GHz & 2.385e3 \\%
        $p_\mathrm{measure}$ & 1.000e-2 & 8.810e-1&&& \\%
        \hline%
        \end{tabular}}%
    \caption{\label{table:parameters_after_simple_optimization}The upper part of the table presents the device and control parameters after the optimization conducted in~\autoref{sec:results_a_simple_gradient_optimization}. The listed gradient corresponds to $\nabla O (\hamparam)$, where $O$ represents the objective function defined in~\autoref{eq:objective_grad_logical_error_rates}. This gradient is an approximation of $\nabla \ler (\hamparam)$, where $\ler$ signifies the logical-$X$ error rate of a $d=7$ surface code.
    The lower part of the table includes the error rates $p^{(j)}_k$ for simultaneous single and two-qubit gates, along with other noise model quantities. The gradients for these values are estimated using the finite-difference method, specifically $\sfrac{\Delta \ler}{\Delta p_k^{(j)}}$. To elaborate, we vary one parameter $p$ at a time with $\Delta p = 0.1 p$, and then compute $\ler$ at $p\pm \Delta p$ by simulating a syndrome circuit for $5\times 10^5$ iterations. The reason for $\sfrac{\partial \ler}{\partial p_1^{(1)}} = 0$ is discussed in~\autoref{sec:qec_sim_workflow}. The gradient units are reciprocals of the corresponding values' units.
    The depolarizing rate in~\autoref{eq:add_decoherence_error_to_unitary} is $r= 10^{-5}$ GHz. Leakage is computed using the formula $p_{\mathrm{leak}}=1-\operatorname{tr}(C^{\dagger}C)/D$, where $C$ is introduced in~\autoref{eq:time_evo_result_and_leakage}, and $D=64$ represents the dimension of $C$. The computed $p_{\mathrm{leak}}$ in this manner corresponds to the average leakage of the 6 qubits.}
\end{table*}

\begin{figure}[htbp]
    \includegraphics[width=\linewidth]{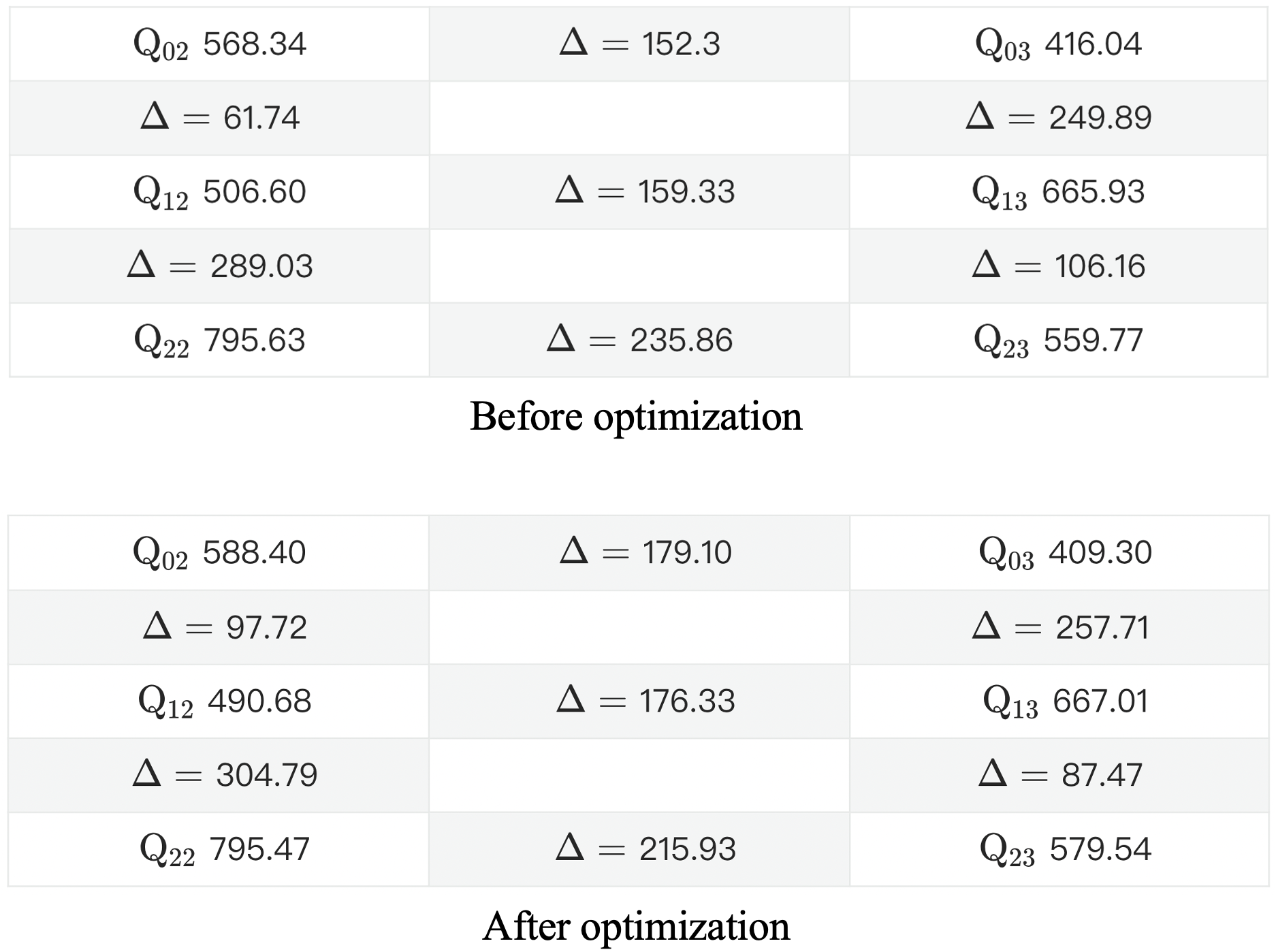}
    \caption{We illustrate the frequencies of the 6 qubits before and after the optimization conducted in~\autoref{sec:results_a_simple_gradient_optimization}.
The units of all values are Mhz.
$\Delta$ is the frequency difference between the two neighboring qubits.
\label{fig:frequency_table}}
\end{figure}

In~\autoref{fig:frequency_table}, we illustrate the frequencies of the 6 qubits before and after the optimization done in~\autoref{sec:results_a_simple_gradient_optimization}.

\normalsize%
\renewcommand{\arraystretch}{1.5}%
\begin{table}
    \centering

\begin{tabular}{| c | c | c | c | c | c | c |}%
\hline%
$\mathrm{Q}_{02}$&$\mathrm{Q}_{03}$&$\mathrm{Q}_{12}$&$\mathrm{Q}_{13}$&$\mathrm{Q}_{22}$&$\mathrm{Q}_{23}$&Error Rate\\%
\hline%
\textcolor{black}{%
I%
}&\textcolor{black}{%
I%
}&\textcolor{black}{%
I%
}&\textcolor{black}{%
I%
}&\textcolor{black}{%
I%
}&\textcolor{black}{%
I%
}&\textcolor{black}{%
9.9190e{-}01%
}\\%
\textcolor{orange}{%
Z%
}&\textcolor{orange}{%
I%
}&\textcolor{orange}{%
Z%
}&\textcolor{orange}{%
I%
}&\textcolor{orange}{%
I%
}&\textcolor{orange}{%
I%
}&\textcolor{orange}{%
1.9426e{-}03%
}\\%
\textcolor{orange}{%
Y%
}&\textcolor{orange}{%
I%
}&\textcolor{orange}{%
Y%
}&\textcolor{orange}{%
I%
}&\textcolor{orange}{%
I%
}&\textcolor{orange}{%
I%
}&\textcolor{orange}{%
1.6123e{-}03%
}\\%
\textcolor{orange}{%
Y%
}&\textcolor{orange}{%
I%
}&\textcolor{orange}{%
Z%
}&\textcolor{orange}{%
I%
}&\textcolor{orange}{%
I%
}&\textcolor{orange}{%
I%
}&\textcolor{orange}{%
1.4277e{-}03%
}\\%
\textcolor{orange}{%
Z%
}&\textcolor{orange}{%
I%
}&\textcolor{orange}{%
Y%
}&\textcolor{orange}{%
I%
}&\textcolor{orange}{%
I%
}&\textcolor{orange}{%
I%
}&\textcolor{orange}{%
1.1817e{-}03%
}\\%
\textcolor{orange}{%
Y%
}&\textcolor{orange}{%
I%
}&\textcolor{orange}{%
X%
}&\textcolor{orange}{%
I%
}&\textcolor{orange}{%
I%
}&\textcolor{orange}{%
I%
}&\textcolor{orange}{%
4.2612e{-}04%
}\\%
\textcolor{orange}{%
Z%
}&\textcolor{orange}{%
I%
}&\textcolor{orange}{%
X%
}&\textcolor{orange}{%
I%
}&\textcolor{orange}{%
I%
}&\textcolor{orange}{%
I%
}&\textcolor{orange}{%
4.0016e{-}04%
}\\%
\textcolor{orange}{%
X%
}&\textcolor{orange}{%
I%
}&\textcolor{orange}{%
Y%
}&\textcolor{orange}{%
I%
}&\textcolor{orange}{%
I%
}&\textcolor{orange}{%
I%
}&\textcolor{orange}{%
1.4200e{-}04%
}\\%
\textcolor{orange}{%
X%
}&\textcolor{orange}{%
I%
}&\textcolor{orange}{%
Z%
}&\textcolor{orange}{%
I%
}&\textcolor{orange}{%
I%
}&\textcolor{orange}{%
I%
}&\textcolor{orange}{%
1.3597e{-}04%
}\\%
\textcolor{purple}{%
I%
}&\textcolor{purple}{%
I%
}&\textcolor{purple}{%
Y%
}&\textcolor{purple}{%
Z%
}&\textcolor{purple}{%
I%
}&\textcolor{purple}{%
Z%
}&\textcolor{purple}{%
6.2225e{-}05%
}\\%
\textcolor{teal}{%
I%
}&\textcolor{teal}{%
I%
}&\textcolor{teal}{%
Y%
}&\textcolor{teal}{%
I%
}&\textcolor{teal}{%
I%
}&\textcolor{teal}{%
I%
}&\textcolor{teal}{%
5.7191e{-}05%
}\\%
\textcolor{purple}{%
I%
}&\textcolor{purple}{%
I%
}&\textcolor{purple}{%
Y%
}&\textcolor{purple}{%
X%
}&\textcolor{purple}{%
I%
}&\textcolor{purple}{%
X%
}&\textcolor{purple}{%
5.6259e{-}05%
}\\%
\textcolor{purple}{%
I%
}&\textcolor{purple}{%
I%
}&\textcolor{purple}{%
Z%
}&\textcolor{purple}{%
Z%
}&\textcolor{purple}{%
I%
}&\textcolor{purple}{%
X%
}&\textcolor{purple}{%
4.2050e{-}05%
}\\%
\textcolor{teal}{%
Y%
}&\textcolor{teal}{%
I%
}&\textcolor{teal}{%
I%
}&\textcolor{teal}{%
I%
}&\textcolor{teal}{%
I%
}&\textcolor{teal}{%
I%
}&\textcolor{teal}{%
3.6702e{-}05%
}\\%
\textcolor{purple}{%
I%
}&\textcolor{purple}{%
I%
}&\textcolor{purple}{%
Y%
}&\textcolor{purple}{%
Y%
}&\textcolor{purple}{%
I%
}&\textcolor{purple}{%
X%
}&\textcolor{purple}{%
3.4281e{-}05%
}\\%
\textcolor{purple}{%
I%
}&\textcolor{purple}{%
I%
}&\textcolor{purple}{%
Z%
}&\textcolor{purple}{%
X%
}&\textcolor{purple}{%
I%
}&\textcolor{purple}{%
Z%
}&\textcolor{purple}{%
3.2387e{-}05%
}\\%
\textcolor{purple}{%
I%
}&\textcolor{purple}{%
I%
}&\textcolor{purple}{%
Z%
}&\textcolor{purple}{%
Z%
}&\textcolor{purple}{%
I%
}&\textcolor{purple}{%
Y%
}&\textcolor{purple}{%
2.8300e{-}05%
}\\%
\textcolor{purple}{%
I%
}&\textcolor{purple}{%
I%
}&\textcolor{purple}{%
Z%
}&\textcolor{purple}{%
Y%
}&\textcolor{purple}{%
I%
}&\textcolor{purple}{%
Z%
}&\textcolor{purple}{%
2.1804e{-}05%
}\\%
\textcolor{orange}{%
X%
}&\textcolor{orange}{%
I%
}&\textcolor{orange}{%
X%
}&\textcolor{orange}{%
I%
}&\textcolor{orange}{%
I%
}&\textcolor{orange}{%
I%
}&\textcolor{orange}{%
1.8770e{-}05%
}\\%
\textcolor{purple}{%
I%
}&\textcolor{purple}{%
I%
}&\textcolor{purple}{%
Y%
}&\textcolor{purple}{%
I%
}&\textcolor{purple}{%
Y%
}&\textcolor{purple}{%
X%
}&\textcolor{purple}{%
1.8639e{-}05%
}\\%
\hline%
\hline%
\multicolumn{6}{| c |}{\textcolor{teal}{%
Average 1-location $p^{(1)}_1$%
}}&\textcolor{teal}{%
1.9843e{-}05%
}\\%
\multicolumn{6}{| c |}{\textcolor{orange}{%
Average 2-location $p^{(1)}_2$%
}}&\textcolor{orange}{%
1.0443e{-}03%
}\\%
\multicolumn{6}{| c |}{\textcolor{purple}{%
Average 3-location $p^{(1)}_3$%
}}&\textcolor{purple}{%
5.2425e{-}05%
}\\%
\hline%
\end{tabular}%
\caption{Pauli errors with largest probabilities for simultaneous X-gate before optimization conducted in~\autoref{sec:results_a_simple_gradient_optimization}.
These Pauli errors are derived from unitary errors through the formula described in~\autoref{eq:error_prob_from_twirling}, and they do not yet include any decoherence errors. The color assigned to each row serves as a visual indicator of the error's weight.
\label{table:simultaneous_x_errors}}
\end{table}

\normalsize%
\renewcommand{\arraystretch}{1.5}%
\begin{table}
    \centering

\begin{tabular}{| c | c | c | c | c | c | c |}%
    \hline%
    $\mathrm{Q}_{02}$&$\mathrm{Q}_{03}$&$\mathrm{Q}_{12}$&$\mathrm{Q}_{13}$&$\mathrm{Q}_{22}$&$\mathrm{Q}_{23}$&Error Rate\\%
    \hline%
    \textcolor{black}{%
    I%
    }&\textcolor{black}{%
    I%
    }&\textcolor{black}{%
    I%
    }&\textcolor{black}{%
    I%
    }&\textcolor{black}{%
    I%
    }&\textcolor{black}{%
    I%
    }&\textcolor{black}{%
    9.9944e{-}01%
    }\\%
    \textcolor{orange}{%
    I%
    }&\textcolor{orange}{%
    I%
    }&\textcolor{orange}{%
    I%
    }&\textcolor{orange}{%
    Z%
    }&\textcolor{orange}{%
    I%
    }&\textcolor{orange}{%
    Z%
    }&\textcolor{orange}{%
    1.2491e{-}04%
    }\\%
    \textcolor{orange}{%
    I%
    }&\textcolor{orange}{%
    I%
    }&\textcolor{orange}{%
    I%
    }&\textcolor{orange}{%
    X%
    }&\textcolor{orange}{%
    I%
    }&\textcolor{orange}{%
    Y%
    }&\textcolor{orange}{%
    7.9677e{-}05%
    }\\%
    \textcolor{orange}{%
    I%
    }&\textcolor{orange}{%
    I%
    }&\textcolor{orange}{%
    I%
    }&\textcolor{orange}{%
    Y%
    }&\textcolor{orange}{%
    I%
    }&\textcolor{orange}{%
    Y%
    }&\textcolor{orange}{%
    3.5984e{-}05%
    }\\%
    \textcolor{orange}{%
    Z%
    }&\textcolor{orange}{%
    I%
    }&\textcolor{orange}{%
    X%
    }&\textcolor{orange}{%
    I%
    }&\textcolor{orange}{%
    I%
    }&\textcolor{orange}{%
    I%
    }&\textcolor{orange}{%
    1.6916e{-}05%
    }\\%
    \textcolor{orange}{%
    X%
    }&\textcolor{orange}{%
    I%
    }&\textcolor{orange}{%
    Z%
    }&\textcolor{orange}{%
    I%
    }&\textcolor{orange}{%
    I%
    }&\textcolor{orange}{%
    I%
    }&\textcolor{orange}{%
    1.6326e{-}05%
    }\\%
    \textcolor{orange}{%
    Z%
    }&\textcolor{orange}{%
    I%
    }&\textcolor{orange}{%
    Z%
    }&\textcolor{orange}{%
    I%
    }&\textcolor{orange}{%
    I%
    }&\textcolor{orange}{%
    I%
    }&\textcolor{orange}{%
    1.4852e{-}05%
    }\\%
    \textcolor{orange}{%
    X%
    }&\textcolor{orange}{%
    I%
    }&\textcolor{orange}{%
    X%
    }&\textcolor{orange}{%
    I%
    }&\textcolor{orange}{%
    I%
    }&\textcolor{orange}{%
    I%
    }&\textcolor{orange}{%
    1.3981e{-}05%
    }\\%
    \textcolor{purple}{%
    X%
    }&\textcolor{purple}{%
    Y%
    }&\textcolor{purple}{%
    Z%
    }&\textcolor{purple}{%
    I%
    }&\textcolor{purple}{%
    I%
    }&\textcolor{purple}{%
    I%
    }&\textcolor{purple}{%
    1.3083e{-}05%
    }\\%
    \textcolor{purple}{%
    Z%
    }&\textcolor{purple}{%
    Z%
    }&\textcolor{purple}{%
    Z%
    }&\textcolor{purple}{%
    I%
    }&\textcolor{purple}{%
    I%
    }&\textcolor{purple}{%
    I%
    }&\textcolor{purple}{%
    1.2957e{-}05%
    }\\%
    \textcolor{purple}{%
    X%
    }&\textcolor{purple}{%
    Z%
    }&\textcolor{purple}{%
    X%
    }&\textcolor{purple}{%
    I%
    }&\textcolor{purple}{%
    I%
    }&\textcolor{purple}{%
    I%
    }&\textcolor{purple}{%
    1.2305e{-}05%
    }\\%
    \textcolor{purple}{%
    Z%
    }&\textcolor{purple}{%
    Y%
    }&\textcolor{purple}{%
    X%
    }&\textcolor{purple}{%
    I%
    }&\textcolor{purple}{%
    I%
    }&\textcolor{purple}{%
    I%
    }&\textcolor{purple}{%
    1.1378e{-}05%
    }\\%
    \textcolor{purple}{%
    Z%
    }&\textcolor{purple}{%
    I%
    }&\textcolor{purple}{%
    Z%
    }&\textcolor{purple}{%
    Z%
    }&\textcolor{purple}{%
    I%
    }&\textcolor{purple}{%
    I%
    }&\textcolor{purple}{%
    6.4211e{-}06%
    }\\%
    \textcolor{purple}{%
    X%
    }&\textcolor{purple}{%
    I%
    }&\textcolor{purple}{%
    X%
    }&\textcolor{purple}{%
    Z%
    }&\textcolor{purple}{%
    I%
    }&\textcolor{purple}{%
    I%
    }&\textcolor{purple}{%
    6.1497e{-}06%
    }\\%
    \textcolor{purple}{%
    Z%
    }&\textcolor{purple}{%
    I%
    }&\textcolor{purple}{%
    X%
    }&\textcolor{purple}{%
    Y%
    }&\textcolor{purple}{%
    I%
    }&\textcolor{purple}{%
    I%
    }&\textcolor{purple}{%
    4.9756e{-}06%
    }\\%
    \textcolor{red}{%
    Y%
    }&\textcolor{red}{%
    I%
    }&\textcolor{red}{%
    X%
    }&\textcolor{red}{%
    X%
    }&\textcolor{red}{%
    I%
    }&\textcolor{red}{%
    Z%
    }&\textcolor{red}{%
    4.8337e{-}06%
    }\\%
    \textcolor{purple}{%
    X%
    }&\textcolor{purple}{%
    Z%
    }&\textcolor{purple}{%
    Y%
    }&\textcolor{purple}{%
    I%
    }&\textcolor{purple}{%
    I%
    }&\textcolor{purple}{%
    I%
    }&\textcolor{purple}{%
    4.6191e{-}06%
    }\\%
    \textcolor{purple}{%
    I%
    }&\textcolor{purple}{%
    Z%
    }&\textcolor{purple}{%
    Z%
    }&\textcolor{purple}{%
    Z%
    }&\textcolor{purple}{%
    I%
    }&\textcolor{purple}{%
    I%
    }&\textcolor{purple}{%
    4.6046e{-}06%
    }\\%
    \textcolor{purple}{%
    X%
    }&\textcolor{purple}{%
    I%
    }&\textcolor{purple}{%
    Z%
    }&\textcolor{purple}{%
    Y%
    }&\textcolor{purple}{%
    I%
    }&\textcolor{purple}{%
    I%
    }&\textcolor{purple}{%
    4.2224e{-}06%
    }\\%
    \textcolor{purple}{%
    I%
    }&\textcolor{purple}{%
    Y%
    }&\textcolor{purple}{%
    X%
    }&\textcolor{purple}{%
    Z%
    }&\textcolor{purple}{%
    I%
    }&\textcolor{purple}{%
    I%
    }&\textcolor{purple}{%
    4.2033e{-}06%
    }\\%
    \hline%
    \hline%
    \multicolumn{6}{| c |}{\textcolor{teal}{%
    Average 1-location $p^{(1)}_1$ %
    }}&\textcolor{teal}{%
    8.4187e{-}08%
    }\\%
    \multicolumn{6}{| c |}{\textcolor{orange}{%
    Average 2-location $p^{(1)}_2$ %
    }}&\textcolor{orange}{%
    4.4662e{-}05%
    }\\%
    \multicolumn{6}{| c |}{\textcolor{purple}{%
    Average 3-location $p^{(1)}_3$ %
    }}&\textcolor{purple}{%
    1.7060e{-}05%
    }\\%
    \hline%
\end{tabular}%
\caption{Pauli errors with largest probabilities for simultaneous X-gate after optimization conducted in~\autoref{sec:results_a_simple_gradient_optimization}.
This is the Pauli errors converted from the unitary error using~\autoref{eq:error_prob_from_twirling} and we have not added decoherence error yet.
The color of each row serves as a visual indicator of the error's weight.
We observe that there is a weight-4 error highlighted in red in the table.
\label{table:simultaneous_x_errors_after_opt}}
\end{table}

\normalsize%
\renewcommand{\arraystretch}{1.5}%
\begin{table}
    \centering
    \begin{tabular}{| c | c | c | c |}%
        \hline%
        $\mathrm{Q}_{02}$ $\mathrm{Q}_{03}$&$\mathrm{Q}_{12}$ $\mathrm{Q}_{13}$&$\mathrm{Q}_{22}$ $\mathrm{Q}_{23}$&Error Rate\\%
        \hline%
        \textcolor{black}{%
        II%
        }&\textcolor{black}{%
        II%
        }&\textcolor{black}{%
        II%
        }&\textcolor{black}{%
        9.9880e{-}01%
        }\\%
        \textcolor{orange}{%
        ZZ%
        }&\textcolor{orange}{%
        ZI%
        }&\textcolor{orange}{%
        II%
        }&\textcolor{orange}{%
        2.7328e{-}04%
        }\\%
        \textcolor{orange}{%
        IY%
        }&\textcolor{orange}{%
        ZI%
        }&\textcolor{orange}{%
        II%
        }&\textcolor{orange}{%
        2.6971e{-}04%
        }\\%
        \textcolor{orange}{%
        ZI%
        }&\textcolor{orange}{%
        IX%
        }&\textcolor{orange}{%
        II%
        }&\textcolor{orange}{%
        1.5224e{-}04%
        }\\%
        \textcolor{orange}{%
        ZI%
        }&\textcolor{orange}{%
        ZZ%
        }&\textcolor{orange}{%
        II%
        }&\textcolor{orange}{%
        3.3287e{-}05%
        }\\%
        \textcolor{orange}{%
        ZI%
        }&\textcolor{orange}{%
        IY%
        }&\textcolor{orange}{%
        II%
        }&\textcolor{orange}{%
        3.3147e{-}05%
        }\\%
        \textcolor{orange}{%
        II%
        }&\textcolor{orange}{%
        ZZ%
        }&\textcolor{orange}{%
        ZX%
        }&\textcolor{orange}{%
        1.8533e{-}05%
        }\\%
        \textcolor{orange}{%
        II%
        }&\textcolor{orange}{%
        IY%
        }&\textcolor{orange}{%
        ZX%
        }&\textcolor{orange}{%
        1.8136e{-}05%
        }\\%
        \textcolor{orange}{%
        IZ%
        }&\textcolor{orange}{%
        ZI%
        }&\textcolor{orange}{%
        II%
        }&\textcolor{orange}{%
        1.6007e{-}05%
        }\\%
        \textcolor{purple}{%
        YI%
        }&\textcolor{purple}{%
        ZI%
        }&\textcolor{purple}{%
        ZI%
        }&\textcolor{purple}{%
        1.4841e{-}05%
        }\\%
        \textcolor{purple}{%
        YX%
        }&\textcolor{purple}{%
        ZI%
        }&\textcolor{purple}{%
        ZI%
        }&\textcolor{purple}{%
        1.4332e{-}05%
        }\\%
        \textcolor{orange}{%
        II%
        }&\textcolor{orange}{%
        YI%
        }&\textcolor{orange}{%
        YI%
        }&\textcolor{orange}{%
        1.3722e{-}05%
        }\\%
        \textcolor{orange}{%
        ZY%
        }&\textcolor{orange}{%
        ZI%
        }&\textcolor{orange}{%
        II%
        }&\textcolor{orange}{%
        1.2043e{-}05%
        }\\%
        \textcolor{orange}{%
        ZI%
        }&\textcolor{orange}{%
        IZ%
        }&\textcolor{orange}{%
        II%
        }&\textcolor{orange}{%
        1.1612e{-}05%
        }\\%
        \textcolor{orange}{%
        ZI%
        }&\textcolor{orange}{%
        ZY%
        }&\textcolor{orange}{%
        II%
        }&\textcolor{orange}{%
        1.1557e{-}05%
        }\\%
        \textcolor{purple}{%
        XX%
        }&\textcolor{purple}{%
        ZI%
        }&\textcolor{purple}{%
        ZI%
        }&\textcolor{purple}{%
        9.7003e{-}06%
        }\\%
        \textcolor{orange}{%
        II%
        }&\textcolor{orange}{%
        XI%
        }&\textcolor{orange}{%
        XI%
        }&\textcolor{orange}{%
        8.8906e{-}06%
        }\\%
        \textcolor{purple}{%
        XI%
        }&\textcolor{purple}{%
        ZI%
        }&\textcolor{purple}{%
        ZI%
        }&\textcolor{purple}{%
        7.8611e{-}06%
        }\\%
        \textcolor{orange}{%
        II%
        }&\textcolor{orange}{%
        IZ%
        }&\textcolor{orange}{%
        ZX%
        }&\textcolor{orange}{%
        6.5288e{-}06%
        }\\%
        \textcolor{purple}{%
        YX%
        }&\textcolor{purple}{%
        IY%
        }&\textcolor{purple}{%
        ZY%
        }&\textcolor{purple}{%
        6.3805e{-}06%
        }\\%
        \hline%
        \hline%
        \multicolumn{3}{| c |}{\textcolor{teal}{%
        Average 1-location $p^{(2)}_1$%
        }}&\textcolor{teal}{%
        5.1623e{-}06%
        }\\%
        \multicolumn{3}{| c |}{\textcolor{orange}{%
        Average 2-location $p^{(2)}_2$%
        }}&\textcolor{orange}{%
        5.1148e{-}04%
        }\\%
        \multicolumn{3}{| c |}{\textcolor{purple}{%
        Average 3-location $p^{(2)}_3$%
        }}&\textcolor{purple}{%
        1.3553e{-}04%
        }\\%
        \hline%
    \end{tabular}%
    \caption{Pauli errors with largest probabilities for simultaneous CNOT-gate before optimization conducted in~\autoref{sec:results_a_simple_gradient_optimization}.
This is the Pauli errors converted from the unitary error using~\autoref{eq:error_prob_from_twirling} and we have not added decoherence error yet.
The color of each row serves as a visual indicator of the error's weight.
 \label{table:cnot_errors_origin}}
\end{table}

\normalsize%
\renewcommand{\arraystretch}{1.5}%
\begin{table}
    \centering
    \begin{tabular}{| c | c | c | c |}%
        \hline%
        $\mathrm{Q}_{02}$ $\mathrm{Q}_{03}$&$\mathrm{Q}_{12}$ $\mathrm{Q}_{13}$&$\mathrm{Q}_{22}$ $\mathrm{Q}_{23}$&Error Rate\\%
        \hline%
        \textcolor{black}{%
        II%
        }&\textcolor{black}{%
        II%
        }&\textcolor{black}{%
        II%
        }&\textcolor{black}{%
        9.9934e{-}01%
        }\\%
        \textcolor{orange}{%
        ZI%
        }&\textcolor{orange}{%
        IX%
        }&\textcolor{orange}{%
        II%
        }&\textcolor{orange}{%
        1.0488e{-}04%
        }\\%
        \textcolor{orange}{%
        ZZ%
        }&\textcolor{orange}{%
        ZI%
        }&\textcolor{orange}{%
        II%
        }&\textcolor{orange}{%
        8.6017e{-}05%
        }\\%
        \textcolor{orange}{%
        IY%
        }&\textcolor{orange}{%
        ZI%
        }&\textcolor{orange}{%
        II%
        }&\textcolor{orange}{%
        8.3989e{-}05%
        }\\%
        \textcolor{orange}{%
        IX%
        }&\textcolor{orange}{%
        ZI%
        }&\textcolor{orange}{%
        II%
        }&\textcolor{orange}{%
        4.5332e{-}05%
        }\\%
        \textcolor{purple}{%
        XI%
        }&\textcolor{purple}{%
        ZI%
        }&\textcolor{purple}{%
        ZI%
        }&\textcolor{purple}{%
        3.2373e{-}05%
        }\\%
        \textcolor{purple}{%
        YX%
        }&\textcolor{purple}{%
        ZI%
        }&\textcolor{purple}{%
        ZI%
        }&\textcolor{purple}{%
        1.7822e{-}05%
        }\\%
        \textcolor{purple}{%
        XX%
        }&\textcolor{purple}{%
        ZI%
        }&\textcolor{purple}{%
        ZI%
        }&\textcolor{purple}{%
        1.4132e{-}05%
        }\\%
        \textcolor{orange}{%
        II%
        }&\textcolor{orange}{%
        ZZ%
        }&\textcolor{orange}{%
        ZX%
        }&\textcolor{orange}{%
        1.3416e{-}05%
        }\\%
        \textcolor{orange}{%
        II%
        }&\textcolor{orange}{%
        IY%
        }&\textcolor{orange}{%
        ZX%
        }&\textcolor{orange}{%
        1.3305e{-}05%
        }\\%
        \textcolor{orange}{%
        IZ%
        }&\textcolor{orange}{%
        ZI%
        }&\textcolor{orange}{%
        II%
        }&\textcolor{orange}{%
        7.2254e{-}06%
        }\\%
        \textcolor{orange}{%
        IZ%
        }&\textcolor{orange}{%
        IY%
        }&\textcolor{orange}{%
        II%
        }&\textcolor{orange}{%
        6.2189e{-}06%
        }\\%
        \textcolor{orange}{%
        II%
        }&\textcolor{orange}{%
        IY%
        }&\textcolor{orange}{%
        IY%
        }&\textcolor{orange}{%
        6.0048e{-}06%
        }\\%
        \textcolor{orange}{%
        II%
        }&\textcolor{orange}{%
        ZZ%
        }&\textcolor{orange}{%
        ZZ%
        }&\textcolor{orange}{%
        5.3697e{-}06%
        }\\%
        \textcolor{orange}{%
        ZY%
        }&\textcolor{orange}{%
        ZI%
        }&\textcolor{orange}{%
        II%
        }&\textcolor{orange}{%
        5.2953e{-}06%
        }\\%
        \textcolor{orange}{%
        II%
        }&\textcolor{orange}{%
        ZX%
        }&\textcolor{orange}{%
        IZ%
        }&\textcolor{orange}{%
        4.8373e{-}06%
        }\\%
        \textcolor{orange}{%
        ZX%
        }&\textcolor{orange}{%
        ZZ%
        }&\textcolor{orange}{%
        II%
        }&\textcolor{orange}{%
        4.6782e{-}06%
        }\\%
        \textcolor{orange}{%
        II%
        }&\textcolor{orange}{%
        ZX%
        }&\textcolor{orange}{%
        ZY%
        }&\textcolor{orange}{%
        4.6650e{-}06%
        }\\%
        \textcolor{orange}{%
        YI%
        }&\textcolor{orange}{%
        XI%
        }&\textcolor{orange}{%
        II%
        }&\textcolor{orange}{%
        4.6247e{-}06%
        }\\%
        \textcolor{orange}{%
        ZX%
        }&\textcolor{orange}{%
        IY%
        }&\textcolor{orange}{%
        II%
        }&\textcolor{orange}{%
        4.5411e{-}06%
        }\\%
        \hline%
        \hline%
        \multicolumn{3}{| c |}{\textcolor{teal}{%
        Average 1-location $p^{(2)}_1$%
        }}&\textcolor{teal}{%
        7.3897e{-}06%
        }\\%
        \multicolumn{3}{| c |}{\textcolor{orange}{%
        Average 2-location $p^{(2)}_2$%
        }}&\textcolor{orange}{%
        2.5013e{-}04%
        }\\%
        \multicolumn{3}{| c |}{\textcolor{purple}{%
        Average 3-location $p^{(2)}_3$%
        }}&\textcolor{purple}{%
        1.0883e{-}04%
        }\\%
        \hline%
    \end{tabular}%
    \caption{Pauli errors with largest probabilities for simultaneous CNOT-gate after optimization conducted in~\autoref{sec:results_a_simple_gradient_optimization}.
This is the Pauli errors converted from the unitary error using~\autoref{eq:error_prob_from_twirling} and we have not added decoherence error yet.
The color of each row serves as a visual indicator of the error's weight.
 \label{table:cnot_errors}}
\end{table}

\end{document}